\documentclass[amsmath,reprint,a4paper,prb]{revtex4-1}
\usepackage[final]{graphicx}

\begin{document}

\title{Dynamics of fluids in quenched-random potential energy
  landscapes: A mode-coupling theory approach}

\author{Thomas Konincks}
\author{Vincent Krakoviack}

\affiliation{Univ Lyon, ENS de Lyon, CNRS UMR 5182, Universit{\'e}
  Claude Bernard Lyon 1, Laboratoire de Chimie, F-69342, Lyon,
  France.}

\date{\today}

\begin{abstract}
  Motivated by a number of recent experimental and computational
  studies of the dynamics of fluids plunged in quenched-disordered
  external fields, we report on a theoretical investigation of this
  topic within the framework of the mode-coupling theory, based on the
  simple model of the hard-sphere fluid in a Gaussian random field.
  The possible dynamical arrest scenarios driven by an increase of the
  disorder strength and/or of the fluid density are mapped, and the
  corresponding evolutions of time-dependent quantities typically used
  for the characterization of anomalous self-diffusion are illustrated
  with detailed computations.  Overall, a fairly reasonable picture of
  the dynamics of the system at hand is outlined, which in particular
  involves a non-monotonicity of the self-diffusion coefficient with
  fluid density at fixed disorder strength, in agreement with
  experiments. The disorder correlation length is shown to have a
  strong influence on the latter feature.
\end{abstract}

\maketitle

\section{Introduction}

In the past few years, the effect of a smooth static random potential
energy landscape on the dynamics of a variety of soft matter systems
has been the focus of a number of studies, based on experiments
involving optically generated disordered environments and on computer
simulations of models designed to reproduce such a setup.
\cite{HanDalSchJenEge12SM, HanEge12JPCM, EveZunHanBewLadHeuEge13PRE,
  EveHanZunCapBewDalJenLadHeuCasEge13EPJST, HanSchEge13PRE,
  BewEge16PRA, BewLadPlaZunHeuEge16PCCP, BewSenCapPlaSenEge16JCP,
  Bew16Thesis, ShvRodIzdDesKroKiv10OE, ShvRodIzdLeyDesKroKiv10JO,
  VolVolGig14SR, VolKurCalVolGig14OE, PinVelCalElaGigVolVol16NC,
  PaoLeoAng14JPCM, YokAiz17OLT} The motivations behind these studies
are diverse. From a fundamental point of view, dynamics in
quenched-random environments are an important paradigm of dynamics in
complex systems, often characterized by anomalous relaxation and
transport phenomena. \cite{HavBen87AP, BouGeo90PR, HofFra13RPP} The
possibility to directly monitor particle trajectories and
corresponding extended regimes of subdiffusive motion in colloidal
suspensions exposed to disordered speckle patterns therefore
represents a nice opportunity to gain new insight into this problem.
\cite{HanDalSchJenEge12SM, HanEge12JPCM, EveZunHanBewLadHeuEge13PRE,
  EveHanZunCapBewDalJenLadHeuCasEge13EPJST, HanSchEge13PRE,
  BewEge16PRA, BewLadPlaZunHeuEge16PCCP, BewSenCapPlaSenEge16JCP,
  Bew16Thesis} On the more practical side, interaction of a
soft-matter system with a random light field appears as a convenient
means to tune its dynamical properties and to control its behavior,
\cite{ShvRodIzdDesKroKiv10OE, ShvRodIzdLeyDesKroKiv10JO,
  VolVolGig14SR, VolKurCalVolGig14OE, PinVelCalElaGigVolVol16NC,
  PaoLeoAng14JPCM, YokAiz17OLT} with significant prospects in terms of
applications. Thus, microfluidic speckle sieves and sorters,
displaying the ability to fractionate flowing colloidal mixtures, have
recently been described. \cite{VolVolGig14SR, VolKurCalVolGig14OE} In
the limit of strong interaction with disorder, persistent trapping
occurs, allowing one to manipulate large numbers of particles on which
the imposed optical environment acts as a random array of essentially
inescapable wells. \cite{ShvRodIzdDesKroKiv10OE,
  ShvRodIzdLeyDesKroKiv10JO, YokAiz17OLT}

At the theoretical level, the problem has a quite long history dealing
with independent tracers in random potential energy landscapes. For
Newtonian dynamics, a simple percolation argument has been put
forward, combining a topographic view of the energy landscape with the
constraints of conserved mechanical energy and positive kinetic
energy. \cite{Zim68JPC, ZalSch71PRB, Isi92RMP} In this picture, a
low-energy tracer is always trapped in a finite region of space,
because any allowed initial position is completely surrounded by
impassable barriers. This is not the case at high energy, where an
infinite domain is available, in which a tracer can always meet
saddles it can cross. The border between the two regimes corresponds
to a continuum percolation transition at a well-defined threshold
energy. It drives a diffusion-localization transition in the tracer
motion, in the vicinity of which anomalous diffusion unfolds as a
consequence of the fractal geometry of the incipient percolation
cluster. This scenario has been confirmed by computer simulations of
particles moving in a model speckle field
\cite{PezRobBouBraAllPliAspBouSan11NJP} and manifestly also applies
when the energy landscape is produced by randomly placed obstacles
interacting with the tracers via soft potentials.  \cite{YanZha10JSM,
  SkiSchAarHorDul13PRL, SchSpaHofFraHor15SM, SchSkiThoAarHorDul17PRE}
The case of Brownian dynamics requires more elaborate considerations
and more sophisticated approaches.  \cite{Gen75JSP, Zwa88PNAS,
  MasFerGolWic89JSP, ChaBraCha94JCP, DeeCha94JSP, DeaDruHor07JSM,
  TouDea07JPA, DeaTou08JPA} Indeed, because of the presence of the
fluctuating thermal bath, energy barriers can be overcome through
activated hopping processes nonexistent in the energy-conserving
case. A variety of scenarios results, depending on space dimension and
statistics of the random energy landscape.  \cite{DeaDruHor07JSM,
  TouDea07JPA, DeaTou08JPA} In many important cases, the late-time
behavior of the tracer is proved to remain diffusive at all
temperatures or disorder strengths, with a rapid decrease of the
diffusion coefficient and a rapid increase of the time needed to reach
the asymptotic diffusive regime as temperature decreases or disorder
strength increases.  \cite{MasFerGolWic89JSP, DeeCha94JSP,
  DeaDruHor07JSM, DeaTou08JPA, BanBisSekBag14JCP}

In the present study, our aim is to go beyond the restriction to
independent tracers and to report on a theoretical investigation of
the dynamics of fluids plunged in smooth random potential energy
landscapes encompassing the case of finite-density fully interacting
systems.  There are only few schemes allowing one to tackle such an
issue from first principles. Here, we consider the framework of the
mode-coupling theory (MCT), in broad use in studies of slow dynamical
phenomena in strongly correlated fluid systems.  \cite{LesHouches,
  GotzeBook} From its numerous applications in varied situations, a
clear picture of the strengths and weaknesses of MCT has emerged. On
the one hand, the theory has repeatedly been found able to correctly
predict or reproduce nontrivial dynamical evolutions, such as the
reentry phenomenon and onset of logarithmic relaxation in the glassy
dynamics of colloidal suspensions with short-ranged attractions
\cite{FofDawBulSciZacTar02PRE, ZacFofDawBulSciTar02PRE,
  SciTarZac03PRL, PhaEgePusPoo04PRE} or, more recently, the strong
non-monotonic variation of the dynamics of fluids confined in slit
pores.  \cite{LanBotOetHajFraSch10PRL, ManLanGroOetRaaFraVar14NC} On
the other hand, MCT displays a marked trend towards an overestimation
of dynamical slowing-down with increasing static correlations and
typically predicts sharp ergodicity-breaking transitions where only
smooth dynamical crossovers are to be found in the actual behavior of
the systems under study.  \cite{GotSjo92RPP, Got99JPCM} It should be
anticipated that both these positive and negative features of MCT will
show up in this work as well.

In practice, the present investigation builds on previous MCT studies
of fluids in quenched-disordered environments, that dealt with fluids
confined in random arrays of immobile obstacles representing
streamlined models of amorphous porous solids.  Early work in the
field addressed the single-particle limit, the so-called random
Lorentz gas, in particular, \cite{GotLeuYip81aPRA, GotLeuYip81bPRA,
  Leu83aPRA, Sza04EL} then an extension to finite fluid densities
appeared more recently.  \cite{Kra05PRL, Kra05JPCM, Kra07PRE,
  Kra09PRE, Kra11PRE} The corresponding theoretical predictions have
subsequently been compared with computer simulation results and shown
to outline a very reasonable picture of the dynamics of the systems at
hand, \cite{KurCosKah09PRL, KurCosKah10PRE, KurCosKah11JPCM,
  KimMiySai09EL, KimMiySai10EPJST, KimMiySai11JPCM,
  SpaSchHofVoiFra13SM} thereby supporting the idea that the MCT
framework can indeed be put to good use in studies of fluids in
randomness.

The connection with the present problem follows from the observation
that the MCT equations for the random fluid-matrix systems eventually
make no reference at all to the particulate character of the disorder.
\cite{Kra07PRE, Kra09PRE} This naturally suggests that they might have
a broader domain of application, an intuition easily confirmed by
rederiving them under the sole minimal assumption that the fluid
evolves in some statistically homogeneous environment of unspecified
nature.  \cite{Kra07PRE} Thus, within MCT, the dynamics of fluids in
disordered matrices and in smooth random fields are described by the
same equations, that become specialized to any given class of system
only through the use of the matching structural information as input
for them.  Since many aspects of the predictions of the theory are
generically determined by the nonlinearities of its equations and the
bifurcation schemes they allow under smooth evolutions of the fluid
structure factors, \cite{LesHouches, GotzeBook} it follows that these
MCT dynamics have a lot in common.  In fact, from the mere structure
of the equations basically results only one overall picture of
dynamics in statistically homogeneous quenched randomness, which
interpolates between ideal glassy dynamics in the bulk and a
diffusion-localization transition scenario at vanishing density and
strong disorder.  Therefore, many predictions of the theory are
interchangeable between different classes of systems.  In previous
work on random fluid-matrix systems, \cite{Kra05PRL, Kra05JPCM,
  Kra07PRE, Kra09PRE, Kra11PRE} the focus was on intermediate
scattering functions, in reflection of the interest in studies of the
glass transition under nanoscale confinement.  \cite{AlcMcK05JPCM,
  AlbCoaDosDudGubRadSli06JPCM, Ric11ARPC} Here, we shall rather direct
the discussion towards signatures of anomalous self-diffusion.
Eventually, the whole picture in both cases is given by the reunion of
these results.  Once this somewhat rigid universal framework is
recognized, one easily conceives to investigate specific features of a
given class of systems with common physical ingredients by studying
particular instances only, thought to be representative and chosen for
their simplicity.  This is the case in the present work, where a
special type of random energy landscape is considered.  There is an
obvious caveat, however, inasmuch as the theory is structurally bound
to fail to capture the whole diversity of the behaviors discussed
above, with their possible nontrivial dependences on disorder
statistics and microdynamics. A similar difficulty has recently been
discussed in the case of the Lorentz gas.
\cite{SpaHofKapMecSchFra16PRL}

The paper is organized as follows. In Sec.~\ref{sec:model}, the simple
model of a fluid in a smooth random potential energy landscape
considered in this work is introduced, together with the theoretical
tools needed for its study. In Sec.~\ref{sec:tracer}, the predictions
of MCT for its dynamics in the single-particle and ideal-gas limits
are reported, while the finite-density regime is addressed in
Sec.~\ref{sec:finite}. Finally, Sec.~\ref{sec:conclusion} is devoted
to discussion and concluding remarks.

\section{Model and theoretical framework}
\label{sec:model}

\subsection{Fluid and random-field models}

A simple model of a fluid plunged in a smooth random potential energy
landscape is generically defined by a potential energy function of the
form
\begin{equation}
  V(\mathbf{r}^N) = \sum_{i=1}^{N-1} \sum_{j=i+1}^N
  v(|\mathbf{r}_i - \mathbf{r}_j|) + \sum_{i=1}^N
  u_\text{dis}(\mathbf{r}_i),
\end{equation}
where $N$ is the fluid particle number and $\mathbf{r}^N \equiv
\{\mathbf{r}_1, \mathbf{r}_2, \ldots, \mathbf{r}_N\}$ denotes the
particle position vectors. The first contribution is the usual
interparticle potential energy, expressed as a sum of pair terms
involving a spherically symmetric interaction potential $v(r)$. The
second one corresponds to the random external potential and is a sum
of identical one-body terms, with $u_\text{dis}$ a smooth energy
profile sampled from a functional ensemble equipped with some
predefined probability density. Because of this field probability
distribution, the statistical mechanics of such a system is based on
two types of averages, as is customary in the presence of quenched
disorder. First, one gets the usual average over thermal fluctuations,
denoted by $\langle\cdots\rangle$ and taken for one given
realization of $u_\text{dis}$, then comes the disorder average,
denoted by $\overline{\cdots}$ and taken over all realizations of
$u_\text{dis}$.

In the following, we wish to take advantage of the fact mentioned in
introduction, that the MCT equations remain the same regardless of the
details of the disorder, to keep the model as simple as
possible. Therefore, $v(r)$ is taken as a hard-core potential of range
$\sigma$. As for $u_\text{dis}$, the simplest choice certainly is a
statistically homogeneous Gaussian random field (GRF),
\cite{MenDas94PRL, ThaDasFei00EL} which can be realized in practice in
many different ways, e.g., as a linear superposition of random
sinusoidal waves, \cite{Kra76JFM} as a sum of interactions with
randomly placed impurities, \cite{LifGrePasbook} or through
coarse-graining of a random field, not necessarily Gaussian in nature,
over large enough regions. \cite{ChuDic98PRB} The latter scheme is
directly relevant to polarizable colloids in speckle patterns.
\cite{HanDalSchJenEge12SM, HanEge12JPCM, EveZunHanBewLadHeuEge13PRE,
  EveHanZunCapBewDalJenLadHeuCasEge13EPJST, HanSchEge13PRE,
  BewEge16PRA, BewLadPlaZunHeuEge16PCCP, BewSenCapPlaSenEge16JCP,
  Bew16Thesis} Such a GRF is fully specified by its mean, which can be
set to zero without loss of generality, and its covariance function
$K(r)$, hence
\begin{equation}
  \overline{u_\text{dis}(\mathbf{r})} = 0, \qquad
  \overline{u_\text{dis}(\mathbf{r}) u_\text{dis}(\mathbf{r'})} =
  K(|\mathbf{r}-\mathbf{r'}|).
\end{equation}
Here, for definiteness, a Gaussian covariance is chosen,
\begin{equation}
  K(r) = \epsilon^2 e^{-(r/\xi)^2},
\end{equation}
with $\epsilon$ a measure of the local energy fluctuations and $\xi$
the correlation length of the disorder. 

Finally, the above model can obviously be laid down in any space
dimensionality $d$, in order to reflect the situation in experiments
and computer simulations, where systems in one, two, and three
dimensions have been investigated.  However, in-depth analyses of the
MCT equations derived from first principles for fluids (quantal or
classical) in randomness have shown that a direct quantitative
application of the theory in less than three dimensions gives rise to
serious difficulties.  \cite{Got78SSC, Got79JPC, GotPreWol79SSC,
  Got81PMB, Leu83bPRA, SchHofFraVoi11JPCM} Therefore, all actual
calculations in this work have been performed on the three-dimensional
model, for which the theoretical anomalies are more easily
tamed. \cite{SchHofFraVoi11JPCM} Whether this provides useful insight
into lower-dimensional systems has then to be assessed on qualitative
grounds, based on the genericness and robustness of the bifurcation
singularities ruling the dynamical scenarios of MCT \cite{LesHouches,
  GotzeBook} and on physically motivated expectations for the
evolution of the underlying nonlinear feedback mechanisms with the
parameters of the model.

\subsection{Structural properties}

In order to quantify the structural properties of a fluid in a random
environment and provide input to the MCT equations, a number of static
quantities need to be defined and computed.

In response to a given realization of the random energy landscape, a
fluid develops a complicated inhomogeneous one-body density profile
$\langle \hat\rho^{(1)}(\mathbf{r}) \rangle$, with
$\hat\rho^{(1)}(\mathbf{r})$ denoting the usual one-body density
operator. \cite{macdohansen2ed} Being a functional of a random field,
this profile is a random field itself. Hence, a basic characterization
thereof is in terms of its mean $\overline{\langle
  \hat\rho^{(1)}(\mathbf{r}) \rangle}$, equal to the fluid
number density $\rho$, and of its two-point correlation function,
defining the disconnected total correlation function $h^\text{d}(r)$
through
\begin{equation}
  \rho^2 h^\text{d}(|\mathbf{r}-\mathbf{r'}|) = \overline{\langle
    \hat\rho^{(1)}(\mathbf{r}) \rangle \langle
    \hat\rho^{(1)}(\mathbf{r'}) \rangle } - \rho^2.
\end{equation}
At the two-point level as well, further information on the particle
distribution in the system is gained by directly averaging the
two-body density operator $\hat\rho^{(2)}(\mathbf{r},\mathbf{r'})$,
\cite{macdohansen2ed} leading, as in the bulk, to the total
correlation function $h(r)$ via
\begin{equation}
  \rho^2 h(|\mathbf{r}-\mathbf{r'}|) = \overline{\langle
    \hat\rho^{(2)}(\mathbf{r},\mathbf{r'}) \rangle} -
  \rho^2.
\end{equation}
By mere subtraction, the connected total correlation function
$h^\text{c}(r)$ is obtained,
\begin{equation}
  \rho^2 h^\text{c}(|\mathbf{r}-\mathbf{r'}|) = \overline{\langle
    \hat\rho^{(2)}(\mathbf{r},\mathbf{r'}) \rangle} -
  \overline{\langle \hat\rho^{(1)}(\mathbf{r}) \rangle
    \langle \hat\rho^{(1)}(\mathbf{r'}) \rangle }.
\end{equation}
Correlatively, in reciprocal space, static structure factors of three
types can be formed. With $n_\mathbf{q} = \sum_{j=1}^{N} e^{i
  \mathbf{q} \cdot \mathbf{r}_j}$ a Fourier component of the
microscopic fluid density, $\delta n_\mathbf{q} = n_\mathbf{q} -
\langle n_\mathbf{q} \rangle$, and $\tilde{f}_q$ the
Fourier transform of $f(r)$, they read
\begin{gather}
  S^\text{d}_q = \frac{\overline{ \langle n_\mathbf{q} \rangle \langle
      n_\mathbf{-q} \rangle }}{N} = \rho \tilde{h}^\text{d}_q, \\
  S_q = \frac{\overline{ \langle n_\mathbf{q} n_\mathbf{-q}
      \rangle }}{N} = 1 + \rho \tilde{h}_q,  \\
  S^\text{c}_q = \frac{\overline{ \langle \delta n_\mathbf{q} \delta
      n_\mathbf{-q} \rangle }}{N} = \frac{\overline{ \langle
      n_\mathbf{q} n_\mathbf{-q} \rangle } - \overline{ \langle
      n_\mathbf{q} \rangle \langle n_\mathbf{-q} \rangle }}{N} = 1 +
  \rho \tilde{h}^\text{c}_q,
\end{gather}
following the same nomenclature as above.  The dependence of all the
above functions on a single scalar variable (distance or wavevector
modulus) results from the assumption of statistical isotropy.

The splitting of correlations into connected and disconnected
components, embodied in the identities $h(r) = h^\text{c}(r) +
h^\text{d}(r)$ and $S_q = S^\text{c}_q + S^\text{d}_q$, is a crucial
feature of the physics of quenched-disordered systems.
\cite{GriMaMaz77PRB} It reflects the presence of two qualitatively
different sources of fluctuations in the problem and actually amounts
to the resolution of two contributions endowed with well-defined and
distinct physical contents. Thus, the disconnected correlations
measure the disorder-induced fluctuations of thermal averages, while
the connected ones quantify the thermal fluctuations about these
averages.

In order to compute the above structural quantities, one can resort to
approximate integral equation theories, best derived by combining
standard approaches of liquid-state physics with the replica trick, a
classic tool of the theory of quenched-disordered systems.  One then
proceeds in two steps.  \cite{EdwAnd75JPF,DeaEdw76PT} First, the
nature of the disorder is switched from quenched to annealed, and the
corresponding statistical mechanics of $s$ noninteracting identical
copies (``replicas'') of the disordered system of interest is worked
out for generic integer $s$. Then, the properties of the original
quenched-disordered system are retrieved by taking the limit $s\to 0$,
after an analytic continuation of the latter results has been
performed with due account for the permutational symmetry of the
equivalent replicas.

It turns out that the application of this scheme to a fluid in a GRF
leads to particularly simple results.  \cite{MenDas94PRL,
  ThaDasFei00EL} The key point here is that the annealed random field
in the replicated systems can be straightforwardly integrated out,
hence an exact recasting of the problem in terms of symmetric
$s$-component mixtures with pairwise interactions only (thanks to the
Gaussianity of the disorder),
\begin{equation}\label{eq:effpot}
  v_{ab}(r) = v(r) \delta_{ab} - \beta K(r),
\end{equation}
where $a$ and $b$ denote replica indices ($a,b=1,\ldots,s$),
$\delta_{ab}$ is the usual Kronecker symbol, and $\beta=1/(k_\text{B}
T)$, with $T$ the temperature and $k_\text{B}$ Boltzmann's constant.
The attractive character of the effective disorder-induced interaction
$-\beta K(r)$ reflects the fact that the particles of the replicated
systems tend to cluster, by populating the most favorable locations
and avoiding the least favorable ones in the random potential energy
landscape. From this reformulation, it is immediate that the equations
describing the structure of a fluid in a GRF can be directly derived
from those of the well-established theory of simple fluid mixtures.
\cite{macdohansen2ed} Thus, in the limit $s\to 0$, where the total and
direct correlation functions of the replicated systems obey
$\lim_{s\to 0} h_{ab}(r) = h(r) \delta_{ab} + h^\text{d}(r)
(1-\delta_{ab})$ and $\lim_{s\to 0} c_{ab}(r) = c(r) \delta_{ab} +
c^\text{d}(r) (1-\delta_{ab})$, the Ornstein-Zernike (OZ) equations
for simple mixtures lead to the so-called replica OZ equations,
\begin{gather}
  h(r) = c(r) + \rho (c \otimes h)(r) - \rho (c^\text{d} \otimes
  h^\text{d})(r), \label{OZ1} \\
  h^\text{d}(r) = c^\text{d}(r) + \rho (c \otimes h^\text{d})(r) +
  \rho (c^\text{d} \otimes h)(r) - 2 \rho (c^\text{d} \otimes
  h^\text{d})(r), \label{OZ2}
\end{gather}
where $\otimes$ denotes a convolution in real space and $c(r)$ and
$c^\text{d}(r)$ are the direct correlation function and disconnected
direct correlation function, respectively, from which the connected
direct correlation function $c^\text{c}(r)$ is obtained via the
equality $c(r) = c^\text{c}(r)+c^\text{d}(r)$. The same limiting
process can be applied to approximate closure relations linking the
correlation functions and the interaction potentials. Starting with
the hypernetted-chain (HNC) equations for mixtures,
\cite{macdohansen2ed} this leads to the replica HNC closure,
\begin{gather}
  c(r) = - \beta v(r) + \beta^2 K(r) + h(r) - \ln \left[ 1 +
    h(r) \right], \label{HNC1} \\
  c^\text{d}(r) = \beta^2 K(r) + h^\text{d}(r) - \ln \left[ 1+
    h^\text{d}(r) \right], \label{HNC2}
\end{gather}
that was introduced in previous work \cite{MenDas94PRL, ThaDasFei00EL}
and is used in the following.  The choice of this approximation is the
result of a compromise.  Indeed, as is well known,
\cite{macdohansen2ed} it lacks accuracy in the pure hard-sphere case,
but it performs very well on systems with ultrasoft interactions,
\cite{LanLikWatLow00JPCM, LouBolHan00PRE, KraHanLou03PRE} to which
belongs the inter-replica disorder-induced effective pair potential
$v_{ab}(r) = - \beta K(r)$, $a\neq b$.  In any case, we checked that
our main results undergo quantitative changes only if other standard
approximation schemes are used for the computation of structural
properties.

The numerical solutions of Eqs.~\eqref{OZ1}-\eqref{HNC2} in real and
reciprocal spaces with $d=3$ have been obtained by a standard method,
\cite{macdohansen2ed} based on a discretization of the correlation
functions at range values $r_i=i \,\delta r$ with
$i=0,1,\ldots,2^{12}$ and $\delta r=0.01 \sigma$, a Picard iterative
method with Broyles mixing, and fast Fourier transform cycles to
compute the convolutions in the OZ equations.

\subsection{Mode-coupling theory}

The main outcome of MCT consists of time-evolution equations for
normalized autocorrelation functions of density fluctuations, also
called intermediate scattering functions (ISF).  \cite{LesHouches,
  GotzeBook} Specifically, for a fluid in a quenched-random
environment, \cite{Kra07PRE, Kra09PRE} the collective dynamics is
described by the connected ISF,
\begin{equation}
  \phi^\text{c}_q(t) = \frac{\overline{ \langle \delta n_\mathbf{q}(t)
      \delta n_\mathbf{-q}(0) \rangle }}{N S^\text{c}_q} = 
  \frac{\overline{ \langle n_\mathbf{q}(t) n_\mathbf{-q}(0)
      \rangle } - \overline{ \langle n_\mathbf{q}
      \rangle \langle n_\mathbf{-q} \rangle }}{N S^\text{c}_q},
\end{equation}
and the self dynamics by the self ISF,
\begin{equation}
  \phi^\text{s}_q(t)= \overline{\langle n^\text{s}_\mathbf{q}(t)
    n^\text{s}_\mathbf{-q}(0) \rangle },
\end{equation} 
where $n^\text{s}_\mathbf{q} = e^{i \mathbf{q} \cdot
  \mathbf{r}^\text{s}}$, with $\mathbf{r}^\text{s}$ corresponding to
the position of one chosen fluid particle considered as tagged.
Again, the sole dependence of the ISFs on the wavevector modulus
follows from statistical isotropy.  Note that, from a theoretical
point of view, the reference to the connected ISF is a natural one,
since the dynamics of the system indeed proceeds from thermal
fluctuations, while the disorder-induced correlations are static in
nature. However, in experiments and computer simulations, one would
typically access the full ISF,
\begin{equation} \label{fulldens}
  \phi_q(t) = \frac{\overline{ \langle n_\mathbf{q}(t)
      n_\mathbf{-q}(0) \rangle}}{N S_q} = \frac{S^\text{c}_q
    \phi^\text{c}_q(t) + S^\text{d}_q}{S_q},
\end{equation}
in which the static effect of quenched disorder manifests itself as a
time-persistent contribution, $S^\text{d}_q/S_q$, unrelated to any
possible glassiness in the collective dynamics.  \cite{Kra05JPCM} No
such difficulty exists for the self dynamics, since, contrary to
$\langle n_\mathbf{q} \rangle$, $\langle
n^\text{s}_\mathbf{q} \rangle$ vanishes for any infinite
sample.

As a first step, the theory involves the derivation of generalized
Langevin equations for both types of density correlators via standard
projection operator methods.  In a suitable overdamped limit,
\cite{FraFucGotMaySin97PRE, FucGotMay98PRE} which we adopt for
simplicity, taking advantage of the fact that the predictions of MCT
at long times are essentially independent of the details of short-time
dynamics, \cite{LesHouches, GotzeBook} they read
\begin{gather}
  \label{langevinc} \tau^\text{c}_{q} \dot{\phi}^\text{c}_{q}(t) +
  \phi^\text{c}_{q}(t) + \int_0^t m^\text{c}_q(t-\tau)
  \dot{\phi}^\text{c}_{q}(\tau) d\tau = 0, \\
  \label{langevins} \tau^\text{s}_{q} \dot{\phi}^\text{s}_{q}(t) +
  \phi^\text{s}_{q}(t) + \int_0^t m^\text{s}_q(t-\tau)
  \dot{\phi}^\text{s}_{q}(\tau) d\tau = 0,
\end{gather}
with $\tau^\text{c}_{q} = S^\text{c}_q / (D_0 q^2)$,
$\tau^\text{s}_{q} = 1 / (D_0 q^2)$, $D_0$ the short-time diffusivity,
and the initial conditions $\phi^\text{c}_q(0) = \phi^\text{s}_{q}(0)
= 1$.  Here and in the following, the possibility of hydrodynamic
interactions between the particles is disregarded, although they might
be sizeable in the actual experimental systems of interest.
\cite{HanDalSchJenEge12SM, HanEge12JPCM, EveZunHanBewLadHeuEge13PRE,
  EveHanZunCapBewDalJenLadHeuCasEge13EPJST, HanSchEge13PRE,
  BewEge16PRA, BewLadPlaZunHeuEge16PCCP, BewSenCapPlaSenEge16JCP,
  Bew16Thesis} The functions $m^\text{c}_q(t)$ and $m^\text{s}_q(t)$
are the so-called memory kernels, each expressed as the
autocorrelation function of a suitable random force.

Then, in order to capture the effects of caging and of scattering by
static microscopic heterogeneities on the dynamics of density
fluctuations, the memory kernels are evaluated by projecting the
random forces onto products of density modes, $\delta n_\mathbf{k}
\delta n_\mathbf{q-k}$ and $\delta n_\mathbf{k} \langle n_\mathbf{q-k}
\rangle$ for $m^\text{c}_q(t)$, $n^\text{s}_\mathbf{k}
\delta n_\mathbf{q-k}$ and $n^\text{s}_\mathbf{k} \langle
n_\mathbf{q-k} \rangle$ for $m^\text{s}_q(t)$. After some
algebra, \cite{Kra07PRE, Kra09PRE} one gets $m^\text{c}_q(t) =
F^\text{c}_q[\phi^\text{c}(t)]$ and $m^\text{s}_q(t) =
F^\text{s}_q[\phi^\text{c}(t),\phi^\text{s}(t)]$, with the functionals
\begin{equation}
  F^\text{c}_q[f^\text{c}] = \int \frac{d^d\mathbf{k}}{(2\pi)^d}
  \left[ V^{(2)}_{\mathbf{q},\mathbf{k}} f^\text{c}_{k}
    f^\text{c}_{|\mathbf{q-k}|} + V^{(1)}_{\mathbf{q},\mathbf{k}}
    f^\text{c}_{k} \right], \label{kernelc}
\end{equation}
\begin{equation}
  F^\text{s}_q[f^\text{c},f^\text{s}] = \int
  \frac{d^d\mathbf{k}}{(2\pi)^d} \left[
    v^{(2)}_{\mathbf{q},\mathbf{k}} f^\text{s}_{k}
    f^\text{c}_{|\mathbf{q-k}|} + v^{(1)}_{\mathbf{q},\mathbf{k}}
    f^\text{s}_{k} \right], \label{kernels}
\end{equation}
and vertices
\begin{gather}
V^{(2)}_{\mathbf{q},\mathbf{k}} = \frac{1}{2} \rho S^\text{c}_q
\left[\frac{\mathbf{q}\cdot\mathbf{k}}{q^2} \tilde{c}^\text{c}_k +
\frac{\mathbf{q}\cdot(\mathbf{q-k})}{q^2}
\tilde{c}^\text{c}_{|\mathbf{q-k}|}\right]^2 S^\text{c}_k
S^\text{c}_{|\mathbf{q-k}|}, \label{v2c} \\
V^{(1)}_{\mathbf{q},\mathbf{k}} = \rho S^\text{c}_q \left[
\frac{\mathbf{q}\cdot\mathbf{k}}{q^2} \tilde{c}^\text{c}_k +
\frac{\mathbf{q}\cdot(\mathbf{q-k})}{q^2} \frac{1}{\rho} \right]^2
S^\text{c}_k S^\text{d}_{|\mathbf{q-k}|}, \label{v1c} \\
v^{(2)}_{\mathbf{q},\mathbf{k}} = \rho \left[
  \frac{\mathbf{q}\cdot(\mathbf{q-k})}{q^2} \right]^2 \left(
  \tilde{c}^\text{c}_{|\mathbf{q-k}|} \right)^2
S^\text{c}_{|\mathbf{q-k}|}, \label{v2s} \\
v^{(1)}_{\mathbf{q},\mathbf{k}} = \left[
\frac{\mathbf{q}\cdot(\mathbf{q-k})}{q^2} \right]^2
\tilde{h}^\text{d}_{|\mathbf{q-k}|}. \label{v1s}
\end{gather}

Gathering Eqs.~\eqref{langevinc}-\eqref{v1s}, a closed self-consistent
set of equations is obtained, from which predictions can be made for
the dynamics of a fluid in randomness, based on structural information
only.  As pointed out in introduction, these equations are very
generic, in that they make no reference to the detailed nature of the
disorder and only involve structural properties of the fluid in its
random environment.  It follows that many predictions based on them
are equally generic.  The specialization to fluids in Gaussian random
fields considered in this work can therefore be expected to provide
results representative of a much broader class of systems.

Analysis of the low $q$ regime leads to additional dynamical equations
for two observables playing a central role in most characterizations
of complex dynamics in fluids, the mean-squared displacement $\delta
r^2(t) = \overline{\langle | \mathbf{r}^\text{s}(t) -
  \mathbf{r}^\text{s}(0) |^2 \rangle}$ and the mean-quartic
displacement $\delta r^4(t) = \overline{\langle |
  \mathbf{r}^\text{s}(t) - \mathbf{r}^\text{s}(0)|^4 \rangle}$. Thus,
plugging the expansion 
\begin{equation}
  \phi^\text{s}_{q}(t) = 1 - \frac{q^2}{2d}  \delta r^2(t) + 
  \frac{q^4}{8d(d+2)} \delta r^4(t) + O(q^6)
\end{equation}
into the MCT equations for $\phi^\text{s}_{q}(t)$, one gets through
identifications at the lowest orders in $q$, \cite{FucGotMay98PRE}
\begin{gather} 
  \delta r^2(t) + D_0 \int_0^t m^{(0)}(t-\tau) \delta r^2(\tau) d\tau
  = 2 d D_0 t, \label{eqMSD} \\
  \begin{split}
    \delta r^4(t) + D_0 \int_0^t m^{(0)}(t-\tau) \delta r^4(\tau)
    d\tau = \qquad\qquad\qquad\qquad \\ 2 (d+2) D_0
    \int_0^t \left[ 2 + m^{(2)}(t-\tau) \right] \delta r^2(\tau) d\tau
    , \label{eqMQD}
  \end{split}
\end{gather}
$m^{(0)}(t)=F^{(0)}[\phi^\text{c}(t),\phi^\text{s}(t)]$, and
$m^{(2)}(t)=F^{(2)}[\phi^\text{c}(t),\phi^\text{s}(t)]$, with the
functionals
\begin{gather} 
  F^{(0)}[f^\text{c},f^\text{s}] = \frac{1}{d} \int
  \frac{d^d\mathbf{k}}{(2\pi)^d} k^{2} \left[ \rho
    (\tilde{c}^\text{c}_{k})^2 S^\text{c}_{k} f^\text{c}_{k} +
    \tilde{h}^\text{d}_{k} \right] f^\text{s}_{k}, \label{memMSD} \\
  \begin{split}
    F^{(2)}[f^\text{c},f^\text{s}] = \frac{3}{d (d+2)} \times
    \qquad\qquad\qquad\qquad\qquad\qquad\qquad \\
    \int \frac{d^d\mathbf{k}}{(2\pi)^d} k^{2} \left[ \rho
      (\tilde{c}^\text{c}_{k})^2 S^\text{c}_{k} f^\text{c}_{k} +
      \tilde{h}^\text{d}_{k} \right] \left[ \frac{\partial^2
        f^\text{s}_{k}}{\partial k^2} + \frac{d-1}{3k} \frac{\partial
        f^\text{s}_{k}}{\partial k} \right]. \label{memMQD}
  \end{split}
\end{gather}

The numerical solutions of these equations with $d=3$ have been
achieved using standard algorithms.  \cite{FucGotHofLat91JPCM,
  FraFucGotMaySin97PRE, FucGotMay98PRE} In particular, after moving to
bipolar coordinates, the wavevector integrals in the memory kernels
have been approximated by Riemann sums, based on an equispaced grid of
$300$ values with step size $\delta q = 2\pi/(2^{12}\delta r) \simeq
0.1534/\sigma$ (imposed by the parameters chosen for the structural
calculations), starting at $\delta q/2$. Such a discretization clearly
involves a cutoff of the low $q$ divergence of some of the vertices,
which might look worrying at first sight. In fact, careful analyses of
the effect of such a cutoff on the solutions of MCT equations have
shown that, on the contrary, it is actually useful to eliminate
spurious long-time anomalies originating in this divergence
\cite{Got78SSC, Got79JPC, GotPreWol79SSC, Got81PMB, Leu83bPRA,
  SchHofFraVoi11JPCM}, and that the quantitative uncertainties
introduced by the arbitrariness of its value remain modest provided
the space dimension is three or larger, \cite{SchHofFraVoi11JPCM}
hence the restriction of the present theoretical study to
three-dimensional systems.

\section{Dynamics in the single-tracer and ideal-gas limits}
\label{sec:tracer}

Since it has been the focus of most past theoretical and computational
studies, we first consider the dynamics in the noninteracting regime,
which indistinctly corresponds to the single-tracer limit ($\rho\to0$
for any $\sigma$) and to the ideal-gas limit ($\sigma\to0$ at any
$\rho$). In both cases, the replica HNC closure turns out to be exact,
as can be checked by a direct calculation of the static correlation
functions introduced above. In particular, one uniformly gets
$h^\text{d}(r) = \exp[\beta^2 K(r)] -1$.

As for the MCT equations, using $S^\text{c}_q \to 1$, $S^\text{d}_q =
\rho \tilde{h}^\text{d}_q$, and $\rho \to 0$ or $\tilde{c}^\text{c}_q
\to 0$, one easily finds that the distinction between connected and
self dynamics vanishes, as does the dependence on $\rho$, and that
each memory kernel reduces to its term linear in the ISF, with
$\tilde{h}^\text{d}_q$ as the sole required structural information.
Furthermore, the disorder correlation length $\xi$, which is the only
relevant lengthscale in the considered limit, can be straightforwardly
scaled out, provided it is used as unit of length together with a unit
of time proportional to $\xi^2/D_0$. Here, this Brownian time is
chosen as $t_{\xi} = \xi^2/(2D_0)$, i.e., the time at which the extent
of free diffusion in one direction of space equals the disorder
correlation length. As expected on physical grounds, this eventually
leaves the relative disorder strength $\Delta=(\beta \epsilon)^2$,
which compares the local potential energy fluctuations to the thermal
energy, as the sole dimensionless parameter controlling the dynamics
of the system.

\begin{figure}
  \centering
  \includegraphics[width=0.4\textwidth,clip]{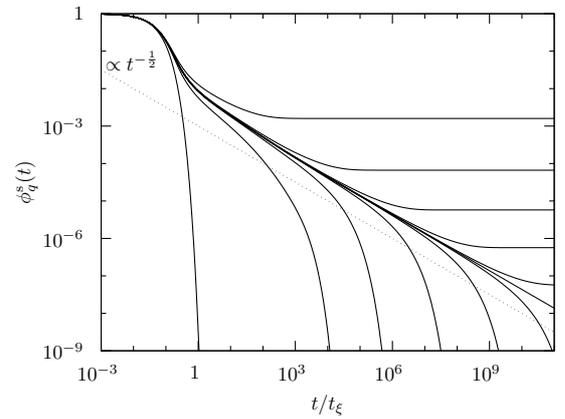}
  \caption{Effect of the relative disorder strength $\Delta$ on the
    time evolution of the intermediate scattering functions
    $\phi^\text{s}_{q}(t)=\phi^\text{c}_{q}(t)$ at $q\simeq 2\pi/\xi$,
    in the noninteracting limit, with $d=3$. From left to right,
    bottom to top: $\Delta=0$, $0.9\Delta_\text{c}$,
    $0.99\Delta_\text{c}$, $0.999\Delta_\text{c}$,
    $0.9999\Delta_\text{c}$, $0.99999\Delta_\text{c}$,
    $\Delta_\text{c}$, $1.00001\Delta_\text{c}$,
    $1.0001\Delta_\text{c}$, $1.001\Delta_\text{c}$,
    $1.01\Delta_\text{c}$, $1.1\Delta_\text{c}$. The dotted line is a
    guide for the eye illustrating the critical decay law
    $\phi^\text{s}_{q}(t) \propto t^{-1/2}$.}
  \label{fig:correlzero}
\end{figure}

The influence of $\Delta$ on the time dependence of the ISFs is
illustrated in Fig.~\ref{fig:correlzero}, where the chosen wavevector
$q \simeq 2\pi/\xi$ corresponds to a real space scale of the order of
$\xi$, but the behavior is essentially the same at any $q$. Increasing
$\Delta$ from zero, the relaxation of the density fluctuations is
expectedly found to slow down, with the gradual development of a weak
algebraic tail extending to longer and longer times, until a critical
value $\Delta_\text{c}\simeq 1.14$ is reached, at which it lasts
forever. Above $\Delta_\text{c}$, the relaxation becomes partially
arrested and an infinite-time plateau appears, whose level
continuously grows with $\Delta$. The decay to this plateau proceeds
through the same algebraic tail as below $\Delta_\text{c}$, which now
progressively recedes as $\Delta$ increases. A standard asymptotic
study of the MCT equations in the vicinity of $\Delta_\text{c}$ gives
$\phi^\text{s}_{q}(t) \propto t^{-1/2}$ for the tail, $\propto
(\Delta-\Delta_\text{c})^{-2}$ for the leading behavior of the
diverging characteristic time marking the end of the tail
symmetrically on both sides of $\Delta_\text{c}$, and $\propto
(\Delta-\Delta_\text{c})$ for the leading behavior of the plateau
level in the arrested state, \cite{LesHouches, FraGot94JPCM, Kra07PRE,
  Kra09PRE} in full agreement with the numerical results.

\begin{figure}
  \centering
  \includegraphics[width=0.4\textwidth,clip]{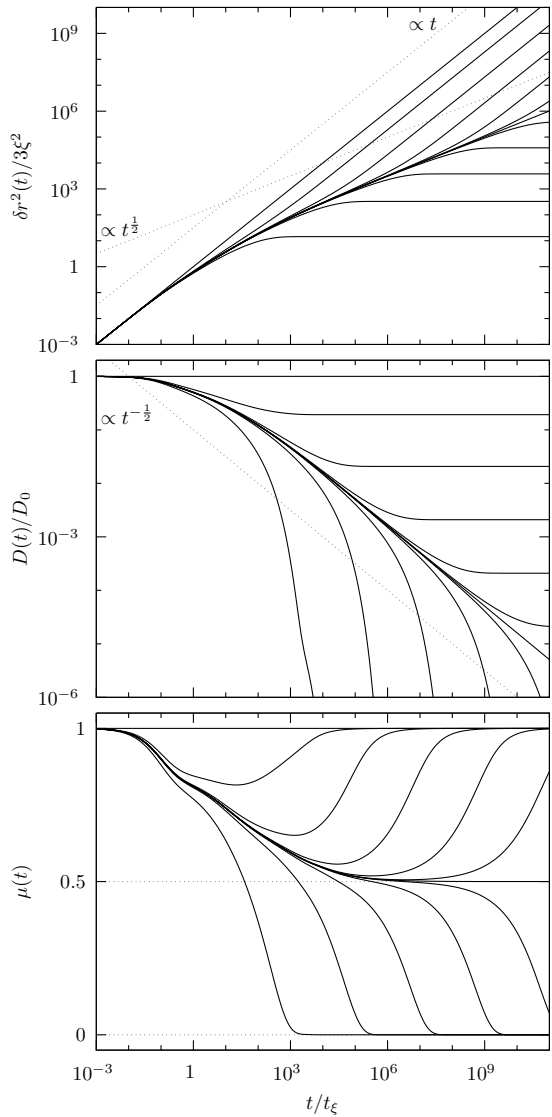}
  \caption{Effect of the relative disorder strength $\Delta$ on the
    time evolution of the normalized mean-squared displacement $\delta
    r^2(t)/(d\xi^2)$ (top panel), of the time-dependent diffusion
    coefficient $D(t)$ (middle panel), and of the local exponent
    $\mu(t)$ (bottom panel), in the noninteracting limit, with
    $d=3$. From top to bottom in all panels: $\Delta=0$,
    $0.9\Delta_\text{c}$, $0.99\Delta_\text{c}$,
    $0.999\Delta_\text{c}$, $0.9999\Delta_\text{c}$,
    $0.99999\Delta_\text{c}$, $\Delta_\text{c}$,
    $1.00001\Delta_\text{c}$, $1.0001\Delta_\text{c}$,
    $1.001\Delta_\text{c}$, $1.01\Delta_\text{c}$,
    $1.1\Delta_\text{c}$. The dotted lines are guides for the eye
    illustrating limiting behaviors.}
  \label{fig:MSDzero}
\end{figure}

From the ISFs, the mean-squared displacement $\delta r^2(t)$ can be
computed via Eqs.~\eqref{eqMSD} and \eqref{memMSD}. The evolution of
its time dependence with $\Delta$ is shown in Fig.~\ref{fig:MSDzero},
together with that of two derived quantities commonly used to magnify
and quantify deviations from normal diffusion, namely, the
time-dependent diffusion coefficient,
\begin{equation}
  D(t) = \frac{1}{2d} \frac{d\delta r^2(t)}{dt}, 
\end{equation}
and the local exponent,
\begin{equation}
  \mu(t) = \frac{d \ln[\delta r^2(t)/(d\xi^2)]}{d\ln(t)}.
\end{equation}
In normal diffusion, $\delta r^2(t)$ is linear in time, hence $D(t)$
is a finite constant and $\mu(t)\equiv 1$. The behavior of the three
functions naturally parallels that of the ISFs. Increasing $\Delta$
between zero and $\Delta_\text{c}$, diffusion progressively slows
down, with the appearance of a subdiffusive regime extending over a
broader and broader time interval (corresponding to the algebraic tail
in the ISFs), after which a normal diffusive behavior is recovered in
the long-time limit. At $\Delta_\text{c}$, the subdiffusive regime
lasts indefinitely. Above $\Delta_\text{c}$, the partially arrested
relaxation of the density fluctuations leads to a similar arrest of
diffusion, with $\delta r^2(t)$ saturating at a finite value, which
decreases with increasing $\Delta$. The same subdiffusive regime as
below $\Delta_\text{c}$ is found, now receding with increasing
$\Delta$. Again, asymptotic results can be derived, which fully agree
with the numerical results and give $\delta r^2(t) \propto t^{1/2}$ in
the subdiffusive regime, a leading-order linear vanishing $\propto
|\Delta-\Delta_\text{c}|$ of the late-time diffusion coefficient
$\lim_{t\to\infty} D(t)$ when $\Delta\to\Delta_\text{c}$ from below,
and a leading order divergence $\propto (\Delta-\Delta_\text{c})^{-1}$
of $\lim_{t\to\infty} \delta r^2(t)$ when $\Delta\to\Delta_\text{c}$
from above. \cite{LesHouches, FraGot94JPCM, Kra07PRE, Kra09PRE}

\begin{figure}[t]
  \centering
  \includegraphics[width=0.4\textwidth,clip]{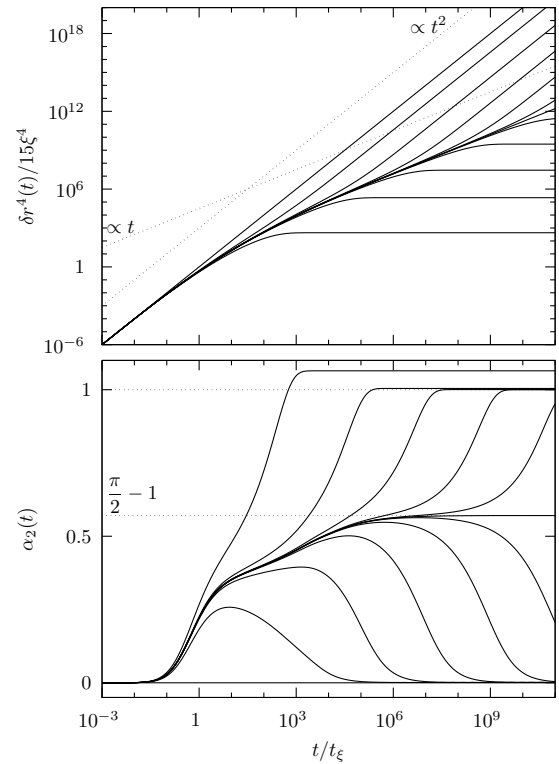}
  \caption{Effect of the relative disorder strength $\Delta$ on the
    time evolution of the normalized mean-quartic displacement $\delta
    r^4(t)/ [d(d+2) \xi^4]$ (top panel) and of the non-Gaussian
    parameter $\alpha_2(t)$ (bottom panel), in the noninteracting
    limit, with $d=3$. From top to bottom in the top panel, from
    bottom to top in the bottom panel: $\Delta=0$,
    $0.9\Delta_\text{c}$, $0.99\Delta_\text{c}$,
    $0.999\Delta_\text{c}$, $0.9999\Delta_\text{c}$,
    $0.99999\Delta_\text{c}$, $\Delta_\text{c}$,
    $1.00001\Delta_\text{c}$, $1.0001\Delta_\text{c}$,
    $1.001\Delta_\text{c}$, $1.01\Delta_\text{c}$,
    $1.1\Delta_\text{c}$. The dotted lines are guides for the eye
    illustrating limiting behaviors.}
  \label{fig:MQDzero}
\end{figure}

The mean-quartic displacement $\delta r^4(t)$ is also a quantity of
interest, since it can be used to form the so-called non-Gaussian
parameter
\begin{equation}
  \alpha_2(t) = \frac{d}{d+2} \frac{\delta r^4(t)}{[\delta r^2(t)]^2}
  - 1, 
\end{equation}
which provides a complementary simple measure of deviations from
normal diffusion, now at the level of the spatial behavior of the
propagator instead of its time dependence. Indeed, in normal
diffusion, the propagator is Gaussian, hence $\alpha_2(t)$ identically
vanishes by construction. The influence of $\Delta$ on the time
dependence of $\delta r^4(t)$ and $\alpha_2(t)$ is displayed in
Fig.~\ref{fig:MQDzero}.  For $\Delta$ between zero and
$\Delta_\text{c}$, the vanishing of $\alpha_2(t)$ at long times
confirms that normal diffusion is recovered in this limit, as inferred
above from the late-time behavior of $\delta r^2(t)$ alone. In the
subdiffusive regime at and around $\Delta_\text{c}$, one gets $\delta
r^4(t)\propto t$, i.e., the power-law behavior one would expect from a
Gaussian propagator with $\delta r^2(t) \propto t^{1/2}$, but a clear
non-Gaussian effect is present in the form of an excess prefactor
[$\alpha_2(t)>0$] asymptotically equal to $\pi/2$. Finally, above
$\Delta_\text{c}$, when diffusion is arrested, $\delta r^4(t)$
saturates at a finite value, with again an excess prefactor with
respect to a Gaussian expectation. This prefactor equals 2
asymptotically close to $\Delta_\text{c}$ and displays a leading-order
linear growth with $\Delta-\Delta_\text{c}$ above.

These MCT predictions reproduce many qualitative features of the
dynamical scenario brought on by an energetic continuum percolation
transition \cite{Zim68JPC, ZalSch71PRB, Isi92RMP} and observed in
various computer simulation studies.
\cite{PezRobBouBraAllPliAspBouSan11NJP, YanZha10JSM,
  SkiSchAarHorDul13PRL, SchSpaHofFraHor15SM, SchSkiThoAarHorDul17PRE}
Indeed, the transition at $\Delta_\text{c}$, with its vanishing of the
late-time diffusivity on one side and divergence of a localization
length on the other, manifestly corresponds to a \emph{bona fide}
diffusion-localization transition, in the vicinity of which an
extended regime of subdiffusive motion unfolds.  Moreover, the local
exponent of the mean-squared displacement and the non-Gaussian
parameter display a reasonable agreement with available simulation
data, in terms of their shapes as functions of time and of their
evolutions with disorder strength.  \cite{SchSpaHofFraHor15SM,
  SchSkiThoAarHorDul17PRE} From a quantitative point of view, however,
an obvious weakness of MCT lies in its predicted critical exponents
and amplitudes, which are constrained to a unique set by the very
structure of its self-consistent equations and the bifurcation scheme
it allows.  On this point, we might incidentally note that the pairing
found in MCT of a subdiffusion exponent $\lim_{t\to\infty} \mu(t)=1/2$
with a critical non-Gaussian parameter $\lim_{t\to\infty}
\alpha_2(t)=\pi/2-1$ also appears as a special case in the
mean-field-like theory of continuous-time random walks (CTRW) with
power-law distributed waiting times, which is \emph{a priori} based on
quite different physical principles.  \cite{HofFra13RPP}

A more subtle discrepancy between the percolation and MCT scenarios is
observed in the diffusive regime.  Indeed, in the percolation picture,
the infinite cluster sustaining diffusive motion always coexists with
bounded regions, in which particles might be indefinitely trapped with
nonvanishing probability.  It follows that, even in the diffusive
regime, the self ISFs and the non-Gaussian parameter never actually
fully relax to zero.  These phenomena have been analyzed in detail in
the case of the Lorentz gas, \cite{KerMet83JPA, KamHofFra08EL,
  FraSpaBauSchHof11JNCS, SpaHofSchMecFra11JPCM} where it could be
shown that the MCT predictions give a fairer description of the
dynamics if assessed against that of the particles confined to the
spanning cluster only.  \cite{SpaSchHofVoiFra13SM} A similar rule of
thumb probably applies to the present class of systems as well.

Although the agreement might be largely coincidental, we should also
mention that the MCT prediction for the transition threshold appears
very reasonable. Indeed, the average potential energy of a tracer in a
GRF, generically given by $-\beta \epsilon^2$, equals $-1.07 \epsilon$
at $\Delta_\text{c}\simeq 1.14$, a value that happens to lie very
close to $-1.03 \epsilon$, the percolation threshold estimated by
Zallen and Scher from empirical arguments.  \cite{ZalSch71PRB}

Qualitative agreement with the diffusion-localization transition
picture means that MCT faces more serious difficulties when dealing
with systems that show a strong slowing-down of their dynamics, but in
which thermal fluctuations eventually restore a long-time normal
diffusive behavior at any disorder strength. This is typically what
happens with Brownian tracers plunged in many types of smooth random
potential energy landscapes (including Gaussian random fields),
\cite{MasFerGolWic89JSP, DeeCha94JSP, DeaDruHor07JSM} but, notably,
not all.  \cite{TouDea07JPA, DeaTou08JPA} In such a case, the
diffusion-localization transition predicted by MCT is an obvious
artifact of the theory. In many respects, this situation appears
reminiscent of the one with bulk glassforming liquids, where the ideal
glass transition predicted by MCT actually has to be looked upon as an
avoided transition, because of intervening thermally activated
processes not included in the theory.  \cite{GotSjo92RPP, Got99JPCM}
Despite this issue, however, the MCT framework has repeatedly been
found to provide a successful platform for qualitative, sometimes
semiquantitative, studies of complex dynamics in soft matter
systems. This is how it will be applied in the next section, in order
to investigate the effect of density on the dynamics of fluids in
random energy landscapes.

\section{Dynamics at finite density}
\label{sec:finite}

In actual experiments, the particle number density might be low, but
it is always finite. \cite{HanDalSchJenEge12SM, HanEge12JPCM,
  EveZunHanBewLadHeuEge13PRE,
  EveHanZunCapBewDalJenLadHeuCasEge13EPJST, HanSchEge13PRE,
  BewEge16PRA, BewLadPlaZunHeuEge16PCCP, BewSenCapPlaSenEge16JCP,
  Bew16Thesis, ShvRodIzdDesKroKiv10OE, ShvRodIzdLeyDesKroKiv10JO,
  VolVolGig14SR, VolKurCalVolGig14OE} Then, the pair interactions
between the particles are expected to interplay with the slowing and
trapping effect of the random energy landscape and to alter the
dynamics compared to the single-tracer limit, to a degree that
naturally depends on the level of crowding in the system. Thus, there
is a clear interest in investigating the influence of density on the
dynamics of fluids in random energy landscapes. In particular, with
applications in mind, such as trapping, sieving, or sorting, which
precisely deal with finite-density systems,
\cite{ShvRodIzdDesKroKiv10OE, ShvRodIzdLeyDesKroKiv10JO,
  VolVolGig14SR, VolKurCalVolGig14OE} such a study might provide
useful guidelines for the choice of optimal operating conditions for
these processes.

As soon as $\rho\neq0$, the particle diameter $\sigma$ is added to the
disorder correlation length $\xi$ as a relevant characteristic
lengthscale, and the ratio $\xi/\sigma$ becomes a defining property of
the system. In the present section, we shall mainly focus on the
typical case $\xi/\sigma=0.5$, which belongs to the most
experimentally relevant regime $\xi/\sigma \lesssim
1$. \cite{HanDalSchJenEge12SM, HanEge12JPCM,
  EveZunHanBewLadHeuEge13PRE,
  EveHanZunCapBewDalJenLadHeuCasEge13EPJST, HanSchEge13PRE,
  BewEge16PRA, BewLadPlaZunHeuEge16PCCP, BewSenCapPlaSenEge16JCP,
  Bew16Thesis} Results for other values of $\xi/\sigma$, also in this
regime, will be mentioned at the end. In the following, $\sigma$ will
be used as unit of length and the corresponding Brownian time
$t_\sigma = \sigma^2/(2D_0)$, i.e., the time it takes a free particle
to diffuse over its own diameter in one direction of space, as unit of
time.

We first consider the dynamical phase diagram of the system, which
represents its state of arrest depending on the reduced density
$\rho\sigma^3$ and relative disorder strength $\Delta=(\beta
\epsilon)^2$.  It is obtained by systematically varying both state
parameters and solving the equations 
\begin{equation}
  \frac{f^\text{c}_q}{1-f^\text{c}_q} = F^\text{c}_q[f^\text{c}],
  \qquad
  \frac{f^\text{s}_q}{1-f^\text{s}_q} =
  F^\text{s}_q[f^\text{c},f^\text{s}],
\end{equation} 
obeyed by the infinite-time limits $f^\text{c}_q=\lim_{t\to\infty}
\phi^\text{c}_{q}(t)$ and $f^\text{s}_q=\lim_{t\to\infty}
\phi^\text{s}_{q}(t)$, \cite{LesHouches, GotzeBook} whose existence
has been proved recently.  \cite{Fra14JPA}

\begin{figure}
  \centering
  \includegraphics[width=0.4\textwidth,clip]{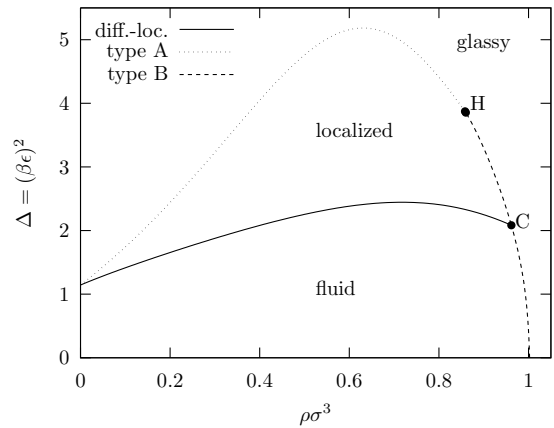}
  \caption{Dynamical phase diagram of a three-dimensional hard-sphere
    fluid plunged in a Gaussian random field with Gaussian covariance,
    for a ratio of the disorder correlation length to the particle
    diameter $\xi/\sigma=0.5$.  The lines correspond to the
    diffusion-localization transitions, type A (continuous) ideal
    glass transitions, and type B (discontinuous) ideal glass
    transitions. H denotes the higher-order singularity formed where
    the two glass transition lines meet and C the crossing point
    between the diffusion-localization and type B glass transition
    lines.}
  \label{fig:phasediag}
\end{figure}

The results are shown in Fig.~\ref{fig:phasediag}. Three dynamical
states are found: (i) fluid, with $f^\text{c}_q=f^\text{s}_q=0$ at all
$q$, in the weak disorder and low-to-moderate density regime, (ii)
glassy, with $f^\text{c}_q\neq0$ and $f^\text{s}_q\neq0$ at all $q$,
in the strong disorder and/or high density regime, and (iii)
localized, with $f^\text{c}_q=0$ and $f^\text{s}_q\neq0$ at all $q$,
in an intermediate regime. The transitions from the fluid or localized
states to the ideal glass are of type B/discontinuous, i.e,
$f^\text{c}_q$ and $f^\text{s}_q$ are discontinuous at the transition,
at high density, and of type A/continuous, i.e, $f^\text{c}_q$ and
$f^\text{s}_q$ are continuous at the transition, at strong
disorder. In the region where the nature of the ideal glass transition
scenario changes, a so-called higher-order singularity is
formed. \cite{LesHouches, GotzeBook} As for the transition from the
fluid to the localized state, it is continuous in terms of
$f^\text{s}_q$ and corresponds to a diffusion-localization transition
essentially similar to the one described in the previous section. At
this point, it might be useful to recall that the current definition
of a glass refers to the existence of dynamically self-induced
time-persistent density correlations \emph{beyond} the equilibrium
static ones imprinted on the fluid by the random field and that,
according to Eq.~\eqref{fulldens}, one always has $\lim_{t\to\infty}
\phi_q(t) = (S^\text{c}_q f^\text{c}_q + S^\text{d}_q)/S_q > 0$ in
quenched disorder, even if $f^\text{c}_q=0$. In particular, this
implies that, although the system is said nonglassy when the
diffusion-localization transition takes place, it does display frozen
disorder-induced amorphous density correlations liable to sustain such
a transition.

In terms of dynamical states and transitions between them, the present
phase diagram is analogous to the one obtained for random fluid-matrix
systems, \cite{Kra07PRE, Kra09PRE} as could actually be anticipated
from the common mathematical structure of the theory for both
problems.  However, details of its shape reflect specificities of the
system at hand.  In this respect, two features seem particularly
relevant.  First, the diffusion-localization transition line shows a
markedly reentrant behavior, with, in the low density regime, an
increase of the localization threshold with increasing density.
Second, the discontinuous glass transition line is quite steep, so
that dynamical arrest at high density appears weakly sensitive to
disorder strength.  For pragmatic reasons, we shall not insist on the
glass transitions at strong disorder, beyond the
diffusion-localization transition line.  Indeed, they lie in a domain
where the fluid might be expected to remain out of equilibrium over
any relevant timescale, because the particles fail to efficiently
redistribute across the system, as entailed by the predicted
diffusion-localization transition.  This creates serious difficulties
in the theory, which is based on equilibrium assumptions, as well as
in experiments and computer simulations, where the system ages.
\cite{BewLadPlaZunHeuEge16PCCP, Bew16Thesis} Therefore, in the
following, we shall make the conservative choice to focus on the sole
dynamical transitions that can be reached directly from the fluid
state.

\begin{figure}[t]
  \centering
  \includegraphics[width=0.4\textwidth,clip]{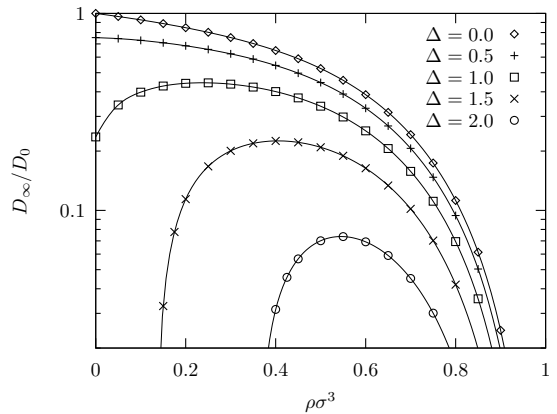}
  \caption{Late-time diffusion coefficient of a three-dimensional
    hard-sphere fluid plunged in a Gaussian random field with Gaussian
    covariance, for a ratio of the disorder correlation length to the
    particle diameter $\xi/\sigma=0.5$. The lines are guides for the
    eye.}
  \label{fig:diffcoeff}
\end{figure}

The diffusion-localization transition is triggered by the increase of
the linear vertices $v^{(1)}_{\mathbf{q},\mathbf{k}}$, which depend on
$\tilde{h}^\text{d}_{q}$ only, as shown by Eq.~\eqref{v1s}.
Accordingly, the shape of the diffusion-localization transition line
and the associated dynamical evolutions within MCT can be
straightforwardly traced back to changes of the latter static
quantity.  At low density, $\tilde{h}^\text{d}_{q}$ essentially
consists of a broad peak at small $q$, whose overall amplitude
decreases with increasing $\rho$ at constant $\Delta$.  The initial
increase of the localization threshold with increasing density
therefore corresponds to the mere compensation of this density effect
by an increase of the relative disorder strength.  At moderate and
high density, $\tilde{h}^\text{d}_{q}$ is small at low $q$ and its
main feature is a peak at a wavevector nearly corresponding to the
contact distance between two particles.  Its height grows with $\rho$
at constant $\Delta$, hence the reversal of the direction of variation
of the localization threshold with density. From the resulting
reentrant behavior of the diffusion-localization transition line, the
prediction by MCT of non-monotonic dynamics at constant $\Delta$ can
be immediately deduced.  Indeed, the isodiffusivity curves, i.e., the
lines in the phase diagram along which the late-time diffusion
coefficient,
\begin{equation}
  D_\infty = \lim_{t\to\infty} D(t) = D_0 \left[ 1+D_0 \int_0^\infty
    m^{(0)}(t) dt \right]^{-1},
\end{equation} 
remains constant, are generically expected to run more or less
parallel to the boundary of the fluid domain, which is a mere limiting
case thereof.  This naturally entails the possibility of non-monotonic
variations of $D_\infty$ with $\rho$ at constant $\Delta$, as easily
confirmed by explicit MCT calculations. Their results are shown in
Fig.~\ref{fig:diffcoeff}, where diffusion coefficients first
increasing, then decreasing, with increasing density at constant
$\Delta$ are readily seen, provided the disorder strength is not too
low.

\begin{figure*}[t]
  \centering
  \includegraphics[width=0.8\textwidth,clip]{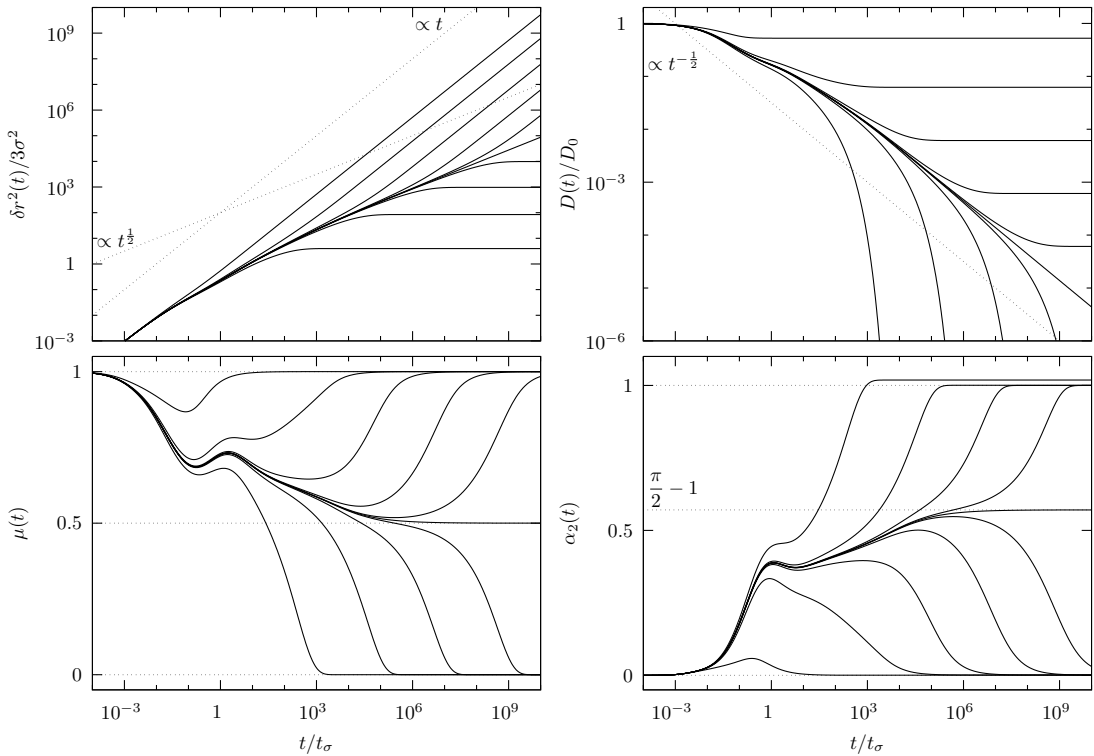}
  \caption{Effect of the relative disorder strength $\Delta$ on the
    time evolution of the normalized mean-squared displacement $\delta
    r^2(t)/(d\sigma^2)$ (top left panel), of the time-dependent
    diffusion coefficient $D(t)$ (top right panel), of the local
    exponent $\mu(t)$ (bottom left panel), and of the non-Gaussian
    parameter $\alpha_2(t)$ (bottom right panel), at density $\rho
    \sigma^3=0.5$, $\xi/\sigma=0.5$, $d=3$. From top to bottom in the
    first three panels, from bottom to top in the last one:
    $\Delta=0$, $0.9\Delta_\text{c}$, $0.99\Delta_\text{c}$,
    $0.999\Delta_\text{c}$, $0.9999\Delta_\text{c}$,
    $\Delta_\text{c}$, $1.0001\Delta_\text{c}$,
    $1.001\Delta_\text{c}$, $1.01\Delta_\text{c}$,
    $1.1\Delta_\text{c}$. The dotted lines are guides for the eye
    illustrating limiting behaviors.}
  \label{fig:rho0.5}
\end{figure*}

In order to gain a mere phenomenological understanding of the behavior
of the system, these observations suggest to look for simple physical
arguments explaining the changes in $\tilde{h}^\text{d}_{q}$, from
which to tentatively infer plausible dynamical evolutions
independently of the MCT formalism.  At low density, the fluid
particles simply tend to occupy the most energetically favorable
regions of the random energy landscape.  But, because of their
excluded-volume interactions, they are progressively forced to occupy
less and less favorable ones as density grows.  On average, at
constant $\Delta$, this leads to a weaker binding of the particles to
their preferred locations and an ensuing reduction of the correlations
imprinted on the fluid by the random field, which are precisely those
measured by $\tilde{h}^\text{d}_{q}$.  At the dynamical level, for the
very same reason, trapping is expected to gradually become less
effective on average with increasing density, with an overall
acceleration of the dynamics as a result.  At moderate and high
density, the shape of $\tilde{h}^\text{d}_{q}$ suggests that, in this
regime, the influence of the random energy landscape on the fluid is
predominantly indirect and mediated by the excluded-volume
interactions.  Thus, a reasonable qualitative picture of the system
seems to be that of a fluid constrained by the particles most strongly
pinned in the wells of the random energy landscape, quite similar to a
fluid-matrix system.  \cite{Kra07PRE, Kra09PRE} In such a case, the
dynamics of the particles whose motion is hindered by those residing
in the lowest energy minima is expected to slow down with increasing
density, by a mere crowding effect.

Clearly, the behavior predicted by MCT is fully consistent with these
phenomenological considerations.  Moreover, experimental evidence has
been gathered in favor of a reentrant dynamics of colloids plunged in
random light fields.  \cite{Bew16Thesis} Therefore, it appears that
MCT correctly captures the dynamical evolutions of the system at hand,
as it does in a number of other cases, \cite{FofDawBulSciZacTar02PRE,
  ZacFofDawBulSciTar02PRE, SciTarZac03PRL, PhaEgePusPoo04PRE,
  LanBotOetHajFraSch10PRL, ManLanGroOetRaaFraVar14NC} despite some
obvious limitations, such as the prediction of sharp transitions that
might actually be avoided.

The steepness of the line of discontinuous ideal glass transitions
delimiting the fluid domain at high density and small-to-moderate
disorder strength can also be understood from simple arguments.  In
this regime, the physics of the system is dominated by excluded-volume
interactions, that efficiently screen the smooth random energy
landscape.  It follows that the disconnected density correlations
remain small compared to the connected ones, that essentially
reproduce the correlations of the dense fluid without random field.
Therefore, the collective quadratic vertices
$V^{(2)}_{\mathbf{q},\mathbf{k}}$, whose increase is then the main
determinant of the discontinuous ideal glass transition, display a
rather modest disorder-strength dependence, leading to the one of the
density at which dynamical arrest takes place.

We might now proceed to report the behavior of some time-dependent
quantities of interest in various regions of the state diagram.  The
intermediate scattering functions have been thoroughly studied in the
case of the fluid-matrix systems, \cite{Kra07PRE, Kra09PRE} and the
generic properties derived there, which actually stem from the overall
structure of the MCT equations, hold without change in the present
case as well.  Therefore, to avoid unnecessary repetition of previous
work, we shall concentrate on quantities pertaining to the
mean-squared and mean-quartic displacements, which are also those most
often reported in experimental and computational studies.

\begin{figure*}[t]
  \centering
  \includegraphics[width=0.8\textwidth,clip]{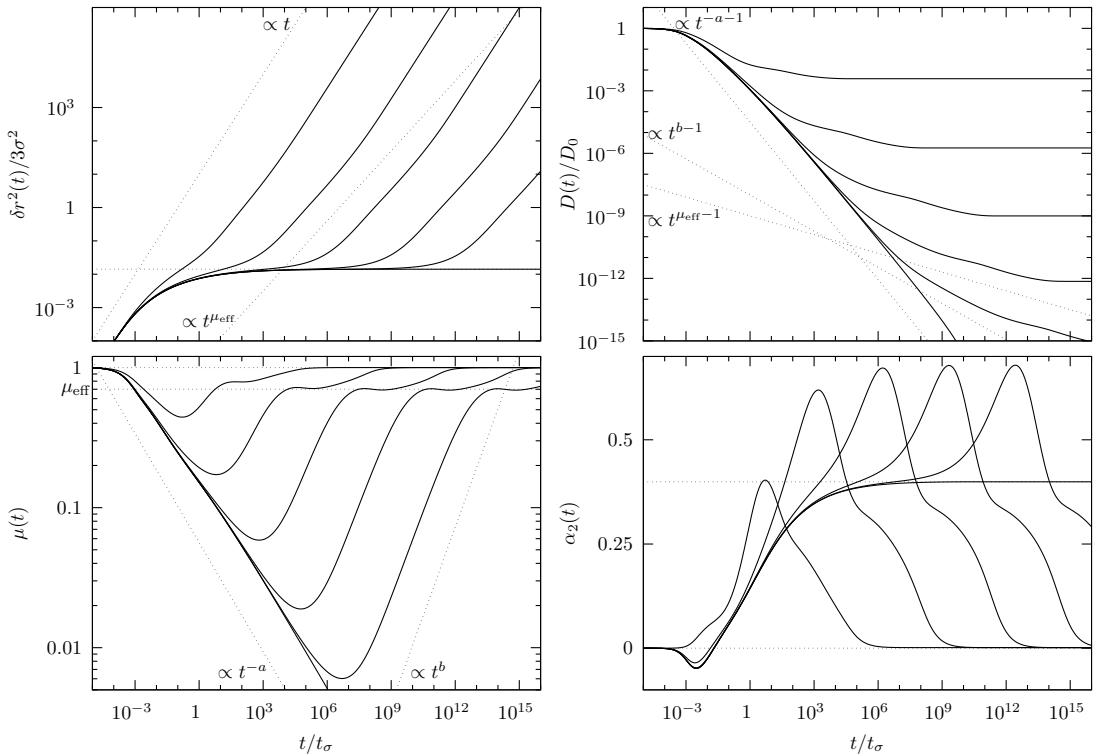}
  \caption{Effect of the fluid number density $\rho\sigma^3$ on the
    time evolution of the normalized mean-squared displacement $\delta
    r^2(t)/(d\sigma^2)$ (top left panel), of the time-dependent
    diffusion coefficient $D(t)$ (top right panel), of the local
    exponent $\mu(t)$ (bottom left panel), and of the non-Gaussian
    parameter $\alpha_2(t)$ (bottom right panel), at relative disorder
    strength $\Delta=2.0$, $\xi/\sigma=0.5$, $d=3$. From top to bottom
    in the first three panels, from left to right in the last one:
    $\rho\sigma^3=0.9\rho_\text{g}\sigma^3$,
    $0.99\rho_\text{g}\sigma^3$, $0.999\rho_\text{g}\sigma^3$,
    $0.9999\rho_\text{g}\sigma^3$, $0.99999\rho_\text{g}\sigma^3$,
    $\rho_\text{g}\sigma^3$. The dotted lines are guides for the eye
    illustrating limiting behaviors.}
  \label{fig:delta2.0}
\end{figure*}

The effect of crossing the diffusion-localization transition line by
increasing $\Delta$ at fixed density is shown in
Fig.~\ref{fig:rho0.5}, where $\rho \sigma^3=0.5$ and the critical
disorder threshold $\Delta_\text{c} \simeq 2.25$.  As already
mentioned, the transition is triggered by the increase of the linear
contribution to the self kernel $m^\text{s}_q(t)$.  Therefore, the
behavior of the reported quantities is essentially the same as in the
noninteracting limit, with alterations in the form of humps at short
times.  These, whose presence here is due to the coupling of the self
and collective dynamics through $v^{(2)}_{\mathbf{q},\mathbf{k}}$ and
whose amplitude increases with density, reflect caging effects readily
seen in the curves at $\Delta=0$.  An important aspect of the
diffusion-localization transition is its absence of counterpart in the
collective dynamics.  Indeed, while the self dynamics becomes complex
and slow or even arrested, the collective one remains simple and fast;
that is, the coherent superposition of the individual anomalous
motions results in an essentially featureless relaxation.  Although
this might seem puzzling at first sight, there is in fact no
particular difficulty or inconsistency here, for a number of reasons.
On the physical side, as already stressed, it is the decay of the
connected ISF which is fast, but the more experimentally relevant full
ISF always displays a strictly arrested contribution corresponding to
the inhomogeneous equilibrium density profile developed by the fluid
in response to its random environment.  Moreover, there is no need in
principle that the particles can diffuse across the system in order to
relax spontaneous thermal density fluctuations about this profile, as
measured by the connected ISF.  Local processes might be sufficient,
and there is no obvious reason why they should develop singularities
because, at very large length- and timescales close to
$\Delta_\text{c}$, the self dynamics might turn out ergodic or not.
Thus, the key point here is the pre-existence of a time-persistent
inhomogeneous density profile, representing a very different situation
compared to the ideal glass transition of a bulk fluid, where the
primary issue is the spontaneous emergence of such a
profile. Remarkably, once it is established, one similarly finds that
it can sustain localization, either right at the transition, as in
one-component systems, \cite{LesHouches, GotzeBook} or after some
delay, in the case of the small component in strongly size-disparate
binary mixtures.  \cite{BosTha87PRL, ThaBos91PRA, ThaBos91PRAb,
  Voi11EPL} In the latter case, the secondary diffusion-localization
transition in the ideal glass is also seen to leave the collective
dynamics unaffected.  At a more formal level, within the memory
function formalism, one can actually show that the collective and self
ISFs are two independent correlation functions, hence they do not need
to display the same type of behavior in general.  \cite{boonyip} This
generic claim is substantiated by studies of simple toy models, in
which one precisely observes self dynamics significantly slower and
more complex than collective dynamics.  \cite{Jep65JMP, DieFulPes80AP,
  Kut81PLA, KehKutBin81PRB} Closer to the present system of interest,
computer simulations of fluids in random obstacle arrays, for which
MCT predicts a similar transition scenario, did evidence strongly
contrasting decay patterns and characteristic timescales for self and
collective dynamics at moderate-to-strong disorder,
\cite{KurCosKah09PRL, KurCosKah10PRE} in qualitative agreement with
the theoretical expectations. \cite{Kra07PRE, Kra09PRE}

The approach towards the discontinuous ideal glass transition line
through an increase of $\rho \sigma^3$ at fixed disorder strength is
illustrated in Fig.~\ref{fig:delta2.0}, with $\Delta=2.0$ and the
ideal glass transition density $\rho_\text{g} \sigma^3 \simeq 0.965$.
In this regime, the relaxation is dominated by the slowing-down of the
collective dynamics, to which the self one is enslaved via the
quadratic term in $m^\text{s}_q(t)$.  Thus, the main observed features
are those familiar and well-studied within the MCT for bulk
glassforming liquids, \cite{LesHouches, GotzeBook,
  FraFucGotMaySin97PRE, FucGotMay98PRE} that we briefly describe.  As
glassy arrest sets in, quantities such as the mean-squared
displacement develop an increasingly extended plateau reflecting
transient localization by caging.  This plateau, to which a diverging
timescale $\propto |\rho - \rho_\text{g}|^{-1/(2a)}$ is attached, is
reached according to a power law $\propto -t^{-a}$, and left according
to another one $\propto t^b$, also known as the von Schweidler law.
At longer times, a simple diffusive behavior is recovered, with a
late-time diffusion coefficient decreasing with density and vanishing
at the transition $\propto |\rho - \rho_\text{g}|^{1/(2a)+1/(2b)}$ to
leading order.  The exponents $a$ and $b$ ($0<a<1/2$ and $b>0$) obey
\begin{equation}
\frac{\Gamma(1-a)^2}{\Gamma(1-2a)} =
\frac{\Gamma(1+b)^2}{\Gamma(1+2b)} = \lambda,  
\end{equation}
where $\Gamma$ denotes Euler's gamma function and the so-called
exponent parameter $\lambda$ is a definite function of the vertices
and non-ergodicity parameters at the transition.  In the standard
scenario valid for bulk fluids, the von Schweidler law directly
crosses over to simple diffusion.  This is also the case here,
provided $\Delta$ is not too large.  Otherwise, if $\Delta$ is such
that the system lies close enough to the diffusion-localization
transition line, a transient domain of subdiffusive motion inserts
itself between the two regimes, as a consequence of the subcritical
but nonnegligible linear term in $m^\text{s}_q(t)$.  The phenomenon is
clearly visible in Fig.~\ref{fig:delta2.0}, where an effective
subdiffusion exponent $\mu_\text{eff}\simeq 0.7$ can be read off,
accompanied by a long time shoulder in $\alpha_2(t)$.

\begin{figure}
  \centering
  \includegraphics[width=0.4\textwidth,clip]{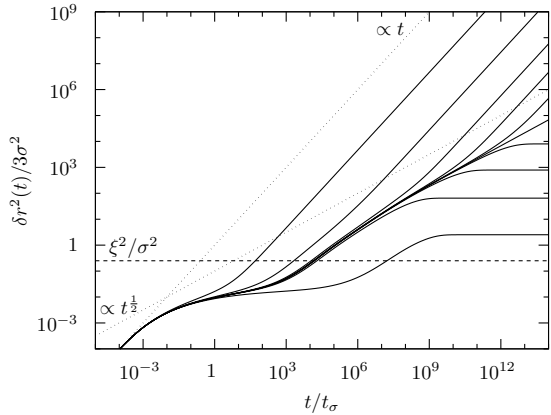}
  \caption{Effect of the relative disorder strength $\Delta$ on the
    time evolution of the normalized mean-squared displacement $\delta
    r^2(t)/(d\sigma^2)$, at density $\rho \sigma^3=0.95$,
    $\xi/\sigma=0.5$, $d=3$. From top to bottom: $\Delta=0$,
    $0.9\Delta_\text{c}$, $0.99\Delta_\text{c}$,
    $0.999\Delta_\text{c}$, $0.9999\Delta_\text{c}$,
    $\Delta_\text{c}$, $1.0001\Delta_\text{c}$,
    $1.001\Delta_\text{c}$, $1.01\Delta_\text{c}$,
    $1.1\Delta_\text{c}$. The dotted lines are guides for the eye
    illustrating limiting behaviors.}
  \label{fig:rho0.95}
\end{figure}

In all cases, the dynamical signatures of caging appear at earlier
times than those of the random energy landscape.  The reason directly
lies in the significant separation of lengthscales between the typical
cage size and the disorder correlation length, as seen in
Fig.~\ref{fig:rho0.95} illustrating the diffusion-localization
transition scenario at high density, with $\rho \sigma^3=0.95$ and
$\Delta_\text{c} \simeq 2.12$.  Clearly, the incipient plateau due to
caging lies well below $\xi^2/\sigma^2$, while subdiffusion starts
above.  Physically, this simply reflects the fact that a particle has
first to escape the cage formed by its neighbors before it can explore
and feel the influence of its quenched-random environment.

\begin{figure*}
  \centering
  \includegraphics[width=0.8\textwidth,clip]{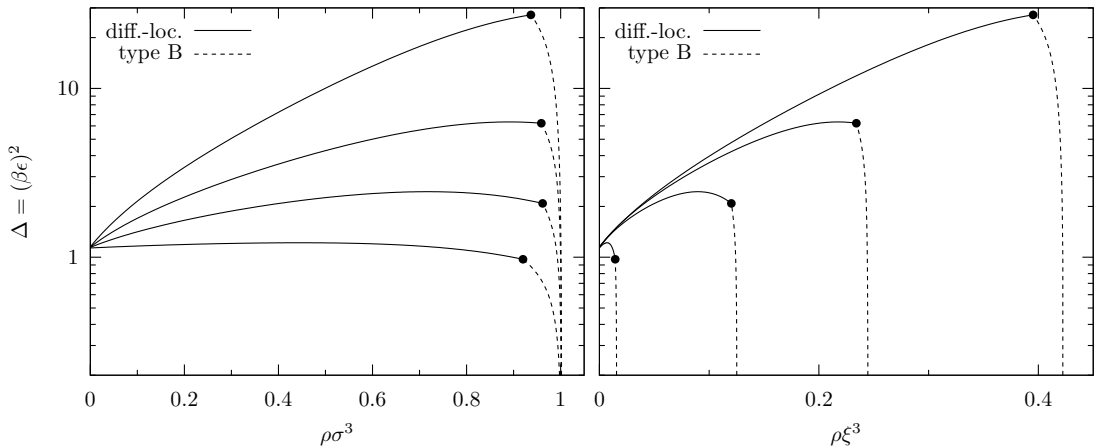}
  \caption{Effect of the ratio of the disorder correlation length to
    the particle diameter $\xi/\sigma$ on the boundaries of the fluid
    state of a three-dimensional hard-sphere fluid plunged in a
    Gaussian random field with Gaussian covariance. Left panel:
    $\Delta$ as a function of $\rho \sigma^3$, right panel: $\Delta$
    as a function of $\rho \xi^3$. The lines correspond to the
    diffusion-localization and type B (discontinuous) ideal glass
    transitions. From bottom to top, left to right: $\xi/\sigma=0.25$,
    $0.5$, $0.625$, $0.75$.}
  \label{fig:xi}
\end{figure*}

We finally consider the effect of the disorder correlation length on
the dynamical phase diagram, by reporting in Fig.~\ref{fig:xi} the
boundaries of the fluid domain at different ratios $\xi/\sigma$.  This
effect is clearly significant, with a fluid domain extending to larger
and larger disorder strengths at high density as $\xi/\sigma$
increases (note the use of a logarithmic scale on the $\Delta$ axis).
To a large extent, this result can be understood from the initial
growth of the localization threshold with density, which is found to
be merely controlled by $\rho \xi^3$, hence the large differences in
initial slope in the $\rho\sigma^3$-$\Delta$ plane.  This is
consistent with the physical picture introduced above, in which the
low-density behavior of the system was interpreted in terms of a
progressive filling of the energetically favorable regions of the
random energy landscape.  Indeed, since $\xi$ acts as a similarity
length for the random field, the density of such regions naturally
scales as $\xi^{-d}$ in space dimension $d$, irrespective of the
detailed energetic criteria used for their definition.  For not too
large values of $\xi/\sigma$, the localization threshold reaches a
maximum at a density increasing with $\xi/\sigma$, beyond which it
becomes a decreasing function of $\rho$.  This evolution has been
explained above as a transition towards the behavior of a fluid
effectively confined by the particles most strongly pinned in the
random energy landscape, the maximum resulting from the complex
interplay of quenched energetic disorder and excluded-volume
interactions.  In such a scenario, since the density of deeply trapped
particles should also scale as $\xi^{-d}$, the effective confinement
strength is expected to decrease with increasing $\xi/\sigma$, hence a
change in the direction of variation of $\Delta_\text{c}$ occurring at
progressively higher density, as indeed predicted by MCT.

\section{Summary and conclusion}
\label{sec:conclusion}

In this paper, the dynamics of fluids plunged in smooth
quenched-random potential energy landscapes has been investigated
within the framework of the mode-coupling theory, based on the simple
model of the hard-sphere fluid in a statistically homogeneous Gaussian
random field.  The possible transitions to dynamically arrested states
triggered by variations of the disorder strength and/or of the fluid
density have been mapped, and the corresponding evolutions of
experimentally relevant time-dependent observables and derived
quantities characterized.

The quality of the MCT predictions for the system at hand seems to be
on a par with what is typically found in applications of this scheme.
On the one hand, the theory inescapably outputs dynamical transitions
as a consequence of increasing static correlations, despite the fact
that this possibility is clearly ruled out in some cases.
\cite{MasFerGolWic89JSP, DeeCha94JSP, DeaDruHor07JSM} This means that
a sharp transition is systematically predicted, where only a strong
dynamical slowing-down likely takes place, which manifests itself in
experiments and simulations in the form of non-ergodic behaviors,
ageing phenomena, and ensuing deviations from exact equilibrium
results (for the late-time diffusion coefficient, for instance).
\cite{HanDalSchJenEge12SM, HanEge12JPCM, EveZunHanBewLadHeuEge13PRE,
  EveHanZunCapBewDalJenLadHeuCasEge13EPJST, HanSchEge13PRE,
  BewEge16PRA, BewLadPlaZunHeuEge16PCCP, BewSenCapPlaSenEge16JCP,
  Bew16Thesis} This is exactly the situation met in MCT studies of
bulk glassforming liquids, which nevertheless remain quite successful.
\cite{GotSjo92RPP, Got99JPCM} On the other hand, the theory is found
to capture some nontrivial qualitative evolutions of the dynamics,
such as its non-monotonicity with fluid density at fixed disorder
strength, which has precisely been observed in experiments.
\cite{Bew16Thesis} The predictions for the local exponent of the
mean-squared displacement and for the non-Gaussian parameter in the
disorder-dominated regime are also in fair agreement with available
simulation data. \cite{SchSpaHofFraHor15SM, SchSkiThoAarHorDul17PRE}
Altogether, it therefore appears that, notwithstanding its obvious
limitations, MCT can be put to good use to gain some insight into the
physics of the considered class of systems.  Thus, as shown in the
previous section, the MCT predictions can be used as a guide in order
to devise simple phenomenological arguments allowing one to explain
the overall evolutions of the dynamics in mere physical terms.

As a microscopic first-principle approach, the present theory also
represents a potentially interesting tool to efficiently investigate
how selected changes to the physical characteristics of the system
might impact its dynamical behavior.  In this respect, the effect of
the disorder correlation length has been addressed as an example and
clearly shown to be significant, although it does not seem to have
been much discussed in the available literature.  We believe that this
type of exploratory studies could be quite helpful for the design of
applications.  For instance, in order to develop microfluidic speckle
sieves or sorters \cite{VolVolGig14SR, VolKurCalVolGig14OE} that
evenly work over an as-a-broad-as-possible colloidal density range,
our results clearly suggest that random optical fields with a short
disorder correlation length should be favored, since they precisely
entail a weaker density dependence of particle diffusion.  Further
refined prescriptions would certainly follow from the extension of the
present study to colloidal mixtures, which are by nature the actual
systems on which the considered fractionation processes operate.  The
corresponding theoretical developments are straightforward, but the
associated numerical calculations are much more demanding. They are
therefore left for future work.

\section*{Acknowledgements}

We thank J{\"o}rg Bewerunge and Stefan Egelhaaf (D{\"u}sseldorf,
Germany) for useful discussions.

%%%END OF MAIN TEXT%%%


%\balance


%%%REFERENCES%%%
%\bibliography{../../../Bibtex/abbrev,../../../Bibtex/confinedfluids_theo,../../../Bibtex/confinedfluids_exp,../../../Bibtex/mct} %You need to replace "rsc" on this line with the name of your .bib file

\begin{mcitethebibliography}{102}
\providecommand*{\natexlab}[1]{#1}
\providecommand*{\mciteSetBstSublistMode}[1]{}
\providecommand*{\mciteSetBstMaxWidthForm}[2]{}
\providecommand*{\mciteBstWouldAddEndPuncttrue}
  {\def\EndOfBibitem{\unskip.}}
\providecommand*{\mciteBstWouldAddEndPunctfalse}
  {\let\EndOfBibitem\relax}
\providecommand*{\mciteSetBstMidEndSepPunct}[3]{}
\providecommand*{\mciteSetBstSublistLabelBeginEnd}[3]{}
\providecommand*{\EndOfBibitem}{}
\mciteSetBstSublistMode{f}
\mciteSetBstMaxWidthForm{subitem}
{(\emph{\alph{mcitesubitemcount}})}
\mciteSetBstSublistLabelBeginEnd{\mcitemaxwidthsubitemform\space}
{\relax}{\relax}

\bibitem[Hanes \emph{et~al.}(2012)Hanes, Dalle-Ferrier, Schmiedeberg, Jenkins,
  and Egelhaaf]{HanDalSchJenEge12SM}
R.~D.~L. Hanes, C.~Dalle-Ferrier, M.~Schmiedeberg, M.~C. Jenkins and S.~U.
  Egelhaaf, \emph{Soft Matter}, 2012, \textbf{8}, 2714--2723\relax
\mciteBstWouldAddEndPuncttrue
\mciteSetBstMidEndSepPunct{\mcitedefaultmidpunct}
{\mcitedefaultendpunct}{\mcitedefaultseppunct}\relax
\EndOfBibitem
\bibitem[Hanes and Egelhaaf(2012)]{HanEge12JPCM}
R.~D.~L. Hanes and S.~U. Egelhaaf, \emph{J. Phys.: Condens. Matter}, 2012,
  \textbf{24}, 464116\relax
\mciteBstWouldAddEndPuncttrue
\mciteSetBstMidEndSepPunct{\mcitedefaultmidpunct}
{\mcitedefaultendpunct}{\mcitedefaultseppunct}\relax
\EndOfBibitem
\bibitem[Evers \emph{et~al.}(2013)Evers, Zunke, Hanes, Bewerunge, Ladadwa,
  Heuer, and Egelhaaf]{EveZunHanBewLadHeuEge13PRE}
F.~Evers, C.~Zunke, R.~Hanes, J.~Bewerunge, I.~Ladadwa, A.~Heuer and
  S.~Egelhaaf, \emph{Phys. Rev. E}, 2013, \textbf{88}, 022125\relax
\mciteBstWouldAddEndPuncttrue
\mciteSetBstMidEndSepPunct{\mcitedefaultmidpunct}
{\mcitedefaultendpunct}{\mcitedefaultseppunct}\relax
\EndOfBibitem
\bibitem[Evers \emph{et~al.}(2013)Evers, Hanes, Zunke, Capellmann, Bewerunge,
  Dalle-Ferrier, Jenkins, Ladadwa, Heuer, Casta{\~n}eda-Priego, and
  Egelhaaf]{EveHanZunCapBewDalJenLadHeuCasEge13EPJST}
F.~Evers, R.~Hanes, C.~Zunke, R.~Capellmann, J.~Bewerunge, C.~Dalle-Ferrier,
  M.~Jenkins, I.~Ladadwa, A.~Heuer, R.~Casta{\~n}eda-Priego and S.~Egelhaaf,
  \emph{Eur. Phys. J. Special Topics}, 2013, \textbf{222}, 2995--3009\relax
\mciteBstWouldAddEndPuncttrue
\mciteSetBstMidEndSepPunct{\mcitedefaultmidpunct}
{\mcitedefaultendpunct}{\mcitedefaultseppunct}\relax
\EndOfBibitem
\bibitem[Hanes \emph{et~al.}(2013)Hanes, Schmiedeberg, and
  Egelhaaf]{HanSchEge13PRE}
R.~D.~L. Hanes, M.~Schmiedeberg and S.~U. Egelhaaf, \emph{Phys. Rev. E}, 2013,
  \textbf{88}, 062133\relax
\mciteBstWouldAddEndPuncttrue
\mciteSetBstMidEndSepPunct{\mcitedefaultmidpunct}
{\mcitedefaultendpunct}{\mcitedefaultseppunct}\relax
\EndOfBibitem
\bibitem[Bewerunge and Egelhaaf(2016)]{BewEge16PRA}
J.~Bewerunge and S.~U. Egelhaaf, \emph{Phys. Rev. A}, 2016, \textbf{93},
  013806\relax
\mciteBstWouldAddEndPuncttrue
\mciteSetBstMidEndSepPunct{\mcitedefaultmidpunct}
{\mcitedefaultendpunct}{\mcitedefaultseppunct}\relax
\EndOfBibitem
\bibitem[Bewerunge \emph{et~al.}(2016)Bewerunge, Ladadwa, Platten, Zunke,
  Heuer, and Egelhaaf]{BewLadPlaZunHeuEge16PCCP}
J.~Bewerunge, I.~Ladadwa, F.~Platten, C.~Zunke, A.~Heuer and S.~U. Egelhaaf,
  \emph{Phys. Chem. Chem. Phys.}, 2016, \textbf{18}, 18887--18895\relax
\mciteBstWouldAddEndPuncttrue
\mciteSetBstMidEndSepPunct{\mcitedefaultmidpunct}
{\mcitedefaultendpunct}{\mcitedefaultseppunct}\relax
\EndOfBibitem
\bibitem[Bewerunge \emph{et~al.}(2016)Bewerunge, Sengupta, Capellmann, Platten,
  Sengupta, and Egelhaaf]{BewSenCapPlaSenEge16JCP}
J.~Bewerunge, A.~Sengupta, R.~F. Capellmann, F.~Platten, S.~Sengupta and S.~U.
  Egelhaaf, \emph{J. Chem. Phys.}, 2016, \textbf{145}, 044905\relax
\mciteBstWouldAddEndPuncttrue
\mciteSetBstMidEndSepPunct{\mcitedefaultmidpunct}
{\mcitedefaultendpunct}{\mcitedefaultseppunct}\relax
\EndOfBibitem
\bibitem[Bewerunge(2016)]{Bew16Thesis}
J.~Bewerunge, \emph{Ph.D. thesis}, Heinrich-Heine-Universit{\"a}t
  D{\"u}sseldorf, 2016\relax
\mciteBstWouldAddEndPuncttrue
\mciteSetBstMidEndSepPunct{\mcitedefaultmidpunct}
{\mcitedefaultendpunct}{\mcitedefaultseppunct}\relax
\EndOfBibitem
\bibitem[Shvedov \emph{et~al.}(2010)Shvedov, Rode, Izdebskaya, Desyatnikov,
  Krolikowski, and Kivshar]{ShvRodIzdDesKroKiv10OE}
V.~G. Shvedov, A.~V. Rode, Y.~V. Izdebskaya, A.~S. Desyatnikov, W.~Krolikowski
  and Y.~S. Kivshar, \emph{Opt. Express}, 2010, \textbf{18}, 3137--3142\relax
\mciteBstWouldAddEndPuncttrue
\mciteSetBstMidEndSepPunct{\mcitedefaultmidpunct}
{\mcitedefaultendpunct}{\mcitedefaultseppunct}\relax
\EndOfBibitem
\bibitem[Shvedov \emph{et~al.}(2010)Shvedov, Rode, Izdebskaya, Leykam,
  Desyatnikov, Krolikowski, and Kivshar]{ShvRodIzdLeyDesKroKiv10JO}
V.~Shvedov, A.~Rode, Y.~Izdebskaya, D.~Leykam, A.~S. Desyatnikov,
  W.~Krolikowski and Y.~S. Kivshar, \emph{J. Opt.}, 2010, \textbf{12},
  124003\relax
\mciteBstWouldAddEndPuncttrue
\mciteSetBstMidEndSepPunct{\mcitedefaultmidpunct}
{\mcitedefaultendpunct}{\mcitedefaultseppunct}\relax
\EndOfBibitem
\bibitem[Volpe \emph{et~al.}(2014)Volpe, Volpe, and Gigan]{VolVolGig14SR}
G.~Volpe, G.~Volpe and S.~Gigan, \emph{Sci. Rep.}, 2014, \textbf{4}, 3936\relax
\mciteBstWouldAddEndPuncttrue
\mciteSetBstMidEndSepPunct{\mcitedefaultmidpunct}
{\mcitedefaultendpunct}{\mcitedefaultseppunct}\relax
\EndOfBibitem
\bibitem[Volpe \emph{et~al.}(2014)Volpe, Kurz, Callegari, Volpe, and
  Gigan]{VolKurCalVolGig14OE}
G.~Volpe, L.~Kurz, A.~Callegari, G.~Volpe and S.~Gigan, \emph{Opt. Express},
  2014, \textbf{22}, 18159--18167\relax
\mciteBstWouldAddEndPuncttrue
\mciteSetBstMidEndSepPunct{\mcitedefaultmidpunct}
{\mcitedefaultendpunct}{\mcitedefaultseppunct}\relax
\EndOfBibitem
\bibitem[Pin{\c{c}}e \emph{et~al.}(2016)Pin{\c{c}}e, Velu, Callegari, Elahi,
  Gigan, Volpe, and Volpe]{PinVelCalElaGigVolVol16NC}
E.~Pin{\c{c}}e, S.~K.~P. Velu, A.~Callegari, P.~Elahi, S.~Gigan, G.~Volpe and
  G.~Volpe, \emph{Nat. Commun.}, 2016, \textbf{7}, 10907\relax
\mciteBstWouldAddEndPuncttrue
\mciteSetBstMidEndSepPunct{\mcitedefaultmidpunct}
{\mcitedefaultendpunct}{\mcitedefaultseppunct}\relax
\EndOfBibitem
\bibitem[Paoluzzi \emph{et~al.}(2014)Paoluzzi, Di~Leonardo, , and
  Angelani]{PaoLeoAng14JPCM}
M.~Paoluzzi, R.~Di~Leonardo,  and L.~Angelani, \emph{J. Phys.: Condens.
  Matter}, 2014, \textbf{26}, 375101\relax
\mciteBstWouldAddEndPuncttrue
\mciteSetBstMidEndSepPunct{\mcitedefaultmidpunct}
{\mcitedefaultendpunct}{\mcitedefaultseppunct}\relax
\EndOfBibitem
\bibitem[Yokoi and Aizu(2017)]{YokAiz17OLT}
N.~Yokoi and Y.~Aizu, \emph{Opt. Laser Technol.}, 2017, \textbf{90},
  226--236\relax
\mciteBstWouldAddEndPuncttrue
\mciteSetBstMidEndSepPunct{\mcitedefaultmidpunct}
{\mcitedefaultendpunct}{\mcitedefaultseppunct}\relax
\EndOfBibitem
\bibitem[Havlin and Ben-Avraham(1987)]{HavBen87AP}
S.~Havlin and D.~Ben-Avraham, \emph{Adv. Phys.}, 1987, \textbf{36},
  695--798\relax
\mciteBstWouldAddEndPuncttrue
\mciteSetBstMidEndSepPunct{\mcitedefaultmidpunct}
{\mcitedefaultendpunct}{\mcitedefaultseppunct}\relax
\EndOfBibitem
\bibitem[Bouchaud and Georges(1990)]{BouGeo90PR}
J.-P. Bouchaud and A.~Georges, \emph{Phys. Rep.}, 1990, \textbf{195},
  127--293\relax
\mciteBstWouldAddEndPuncttrue
\mciteSetBstMidEndSepPunct{\mcitedefaultmidpunct}
{\mcitedefaultendpunct}{\mcitedefaultseppunct}\relax
\EndOfBibitem
\bibitem[H{\"o}fling and Franosch(2013)]{HofFra13RPP}
F.~H{\"o}fling and T.~Franosch, \emph{Rep. Prog. Phys.}, 2013, \textbf{76},
  046602\relax
\mciteBstWouldAddEndPuncttrue
\mciteSetBstMidEndSepPunct{\mcitedefaultmidpunct}
{\mcitedefaultendpunct}{\mcitedefaultseppunct}\relax
\EndOfBibitem
\bibitem[Ziman(1968)]{Zim68JPC}
J.~M. Ziman, \emph{J. Phys. C: Solid State Phys.}, 1968, \textbf{1},
  1532--1538\relax
\mciteBstWouldAddEndPuncttrue
\mciteSetBstMidEndSepPunct{\mcitedefaultmidpunct}
{\mcitedefaultendpunct}{\mcitedefaultseppunct}\relax
\EndOfBibitem
\bibitem[Zallen and Scher(1971)]{ZalSch71PRB}
R.~Zallen and H.~Scher, \emph{Phys. Rev. B}, 1971, \textbf{4}, 4471--4479\relax
\mciteBstWouldAddEndPuncttrue
\mciteSetBstMidEndSepPunct{\mcitedefaultmidpunct}
{\mcitedefaultendpunct}{\mcitedefaultseppunct}\relax
\EndOfBibitem
\bibitem[Isichenko(1992)]{Isi92RMP}
M.~B. Isichenko, \emph{Rev. Mod. Phys.}, 1992, \textbf{64}, 961--1043\relax
\mciteBstWouldAddEndPuncttrue
\mciteSetBstMidEndSepPunct{\mcitedefaultmidpunct}
{\mcitedefaultendpunct}{\mcitedefaultseppunct}\relax
\EndOfBibitem
\bibitem[Pezz{\'e} \emph{et~al.}(2011)Pezz{\'e}, de~Saint-Vincent, Bourdel,
  Brantut, Allard, Plisson, Aspect, Bouyer, and
  Sanchez-Palencia]{PezRobBouBraAllPliAspBouSan11NJP}
L.~Pezz{\'e}, M.~R. de~Saint-Vincent, T.~Bourdel, J.-P. Brantut, B.~Allard,
  T.~Plisson, A.~Aspect, P.~Bouyer and L.~Sanchez-Palencia, \emph{New J.
  Phys.}, 2011, \textbf{13}, 095015\relax
\mciteBstWouldAddEndPuncttrue
\mciteSetBstMidEndSepPunct{\mcitedefaultmidpunct}
{\mcitedefaultendpunct}{\mcitedefaultseppunct}\relax
\EndOfBibitem
\bibitem[Yang and Zhao(2010)]{YanZha10JSM}
J.~Yang and H.~Zhao, \emph{J. Stat. Mech.: Theory Exp.}, 2010, \textbf{2010},
  L12001\relax
\mciteBstWouldAddEndPuncttrue
\mciteSetBstMidEndSepPunct{\mcitedefaultmidpunct}
{\mcitedefaultendpunct}{\mcitedefaultseppunct}\relax
\EndOfBibitem
\bibitem[Skinner \emph{et~al.}(2013)Skinner, Schnyder, Aarts, Horbach, and
  Dullens]{SkiSchAarHorDul13PRL}
T.~O.~E. Skinner, S.~K. Schnyder, D.~G. A.~L. Aarts, J.~Horbach and R.~P.~A.
  Dullens, \emph{Phys. Rev. Lett.}, 2013, \textbf{12}, 128301\relax
\mciteBstWouldAddEndPuncttrue
\mciteSetBstMidEndSepPunct{\mcitedefaultmidpunct}
{\mcitedefaultendpunct}{\mcitedefaultseppunct}\relax
\EndOfBibitem
\bibitem[Schnyder \emph{et~al.}(2015)Schnyder, Spanner, H{\"o}fling, Franosch,
  and Horbach]{SchSpaHofFraHor15SM}
S.~K. Schnyder, M.~Spanner, F.~H{\"o}fling, T.~Franosch and J.~Horbach,
  \emph{Soft Matter}, 2015, \textbf{11}, 701--711\relax
\mciteBstWouldAddEndPuncttrue
\mciteSetBstMidEndSepPunct{\mcitedefaultmidpunct}
{\mcitedefaultendpunct}{\mcitedefaultseppunct}\relax
\EndOfBibitem
\bibitem[Schnyder \emph{et~al.}(2017)Schnyder, Skinner, Thorneywork, Aarts,
  Horbach, and Dullens]{SchSkiThoAarHorDul17PRE}
S.~K. Schnyder, T.~O.~E. Skinner, A.~L. Thorneywork, D.~G. A.~L. Aarts,
  J.~Horbach and R.~P.~A. Dullens, \emph{Phys. Rev. E}, 2017, \textbf{95},
  032602\relax
\mciteBstWouldAddEndPuncttrue
\mciteSetBstMidEndSepPunct{\mcitedefaultmidpunct}
{\mcitedefaultendpunct}{\mcitedefaultseppunct}\relax
\EndOfBibitem
\bibitem[De~Gennes(1975)]{Gen75JSP}
P.~G. De~Gennes, \emph{J. Stat. Phys.}, 1975, \textbf{12}, 463--481\relax
\mciteBstWouldAddEndPuncttrue
\mciteSetBstMidEndSepPunct{\mcitedefaultmidpunct}
{\mcitedefaultendpunct}{\mcitedefaultseppunct}\relax
\EndOfBibitem
\bibitem[Zwanzig(1988)]{Zwa88PNAS}
R.~Zwanzig, \emph{Proc. Natl. Acad. Sci. U.S.A.}, 1988, \textbf{85},
  2029--2030\relax
\mciteBstWouldAddEndPuncttrue
\mciteSetBstMidEndSepPunct{\mcitedefaultmidpunct}
{\mcitedefaultendpunct}{\mcitedefaultseppunct}\relax
\EndOfBibitem
\bibitem[De~Masi \emph{et~al.}(1989)De~Masi, Ferrari, Goldstein, and
  Wick]{MasFerGolWic89JSP}
A.~De~Masi, P.~A. Ferrari, S.~Goldstein and W.~D. Wick, \emph{J. Stat. Phys.},
  1989, \textbf{55}, 787--855\relax
\mciteBstWouldAddEndPuncttrue
\mciteSetBstMidEndSepPunct{\mcitedefaultmidpunct}
{\mcitedefaultendpunct}{\mcitedefaultseppunct}\relax
\EndOfBibitem
\bibitem[Chakraborty \emph{et~al.}(1994)Chakraborty, Bratko, and
  Chandler]{ChaBraCha94JCP}
A.~K. Chakraborty, D.~Bratko and D.~Chandler, \emph{J. Chem. Phys.}, 1994,
  \textbf{100}, 1528--1541\relax
\mciteBstWouldAddEndPuncttrue
\mciteSetBstMidEndSepPunct{\mcitedefaultmidpunct}
{\mcitedefaultendpunct}{\mcitedefaultseppunct}\relax
\EndOfBibitem
\bibitem[Deem and Chandler(1994)]{DeeCha94JSP}
M.~W. Deem and D.~Chandler, \emph{J. Stat. Phys.}, 1994, \textbf{76},
  911--927\relax
\mciteBstWouldAddEndPuncttrue
\mciteSetBstMidEndSepPunct{\mcitedefaultmidpunct}
{\mcitedefaultendpunct}{\mcitedefaultseppunct}\relax
\EndOfBibitem
\bibitem[Dean \emph{et~al.}(2007)Dean, Drummond, and Horgan]{DeaDruHor07JSM}
D.~S. Dean, I.~T. Drummond and R.~R. Horgan, \emph{J. Stat. Mech.: Theory
  Exp.}, 2007, \textbf{2007}, P07013\relax
\mciteBstWouldAddEndPuncttrue
\mciteSetBstMidEndSepPunct{\mcitedefaultmidpunct}
{\mcitedefaultendpunct}{\mcitedefaultseppunct}\relax
\EndOfBibitem
\bibitem[Touya and Dean(2007)]{TouDea07JPA}
C.~Touya and D.~S. Dean, \emph{J. Phys. A: Math. Theor.}, 2007, \textbf{40},
  919--934\relax
\mciteBstWouldAddEndPuncttrue
\mciteSetBstMidEndSepPunct{\mcitedefaultmidpunct}
{\mcitedefaultendpunct}{\mcitedefaultseppunct}\relax
\EndOfBibitem
\bibitem[Dean and Touya(2008)]{DeaTou08JPA}
D.~S. Dean and C.~Touya, \emph{J. Phys. A: Math. Theor.}, 2008, \textbf{41},
  335002\relax
\mciteBstWouldAddEndPuncttrue
\mciteSetBstMidEndSepPunct{\mcitedefaultmidpunct}
{\mcitedefaultendpunct}{\mcitedefaultseppunct}\relax
\EndOfBibitem
\bibitem[Banerjee \emph{et~al.}(2014)Banerjee, Biswas, Seki, and
  Bagchi]{BanBisSekBag14JCP}
S.~Banerjee, R.~Biswas, K.~Seki and B.~Bagchi, \emph{J. Chem. Phys.}, 2014,
  \textbf{141}, 124105\relax
\mciteBstWouldAddEndPuncttrue
\mciteSetBstMidEndSepPunct{\mcitedefaultmidpunct}
{\mcitedefaultendpunct}{\mcitedefaultseppunct}\relax
\EndOfBibitem
\bibitem[G{\"o}tze(1991)]{LesHouches}
W.~G{\"o}tze, in \emph{Liquids, freezing and glass transition, Les Houches
  1989}, ed. J.-P. Hansen, D.~Levesque and J.~Zinn-Justin, North Holland,
  Amsterdam, 1991, pp. 287--503\relax
\mciteBstWouldAddEndPuncttrue
\mciteSetBstMidEndSepPunct{\mcitedefaultmidpunct}
{\mcitedefaultendpunct}{\mcitedefaultseppunct}\relax
\EndOfBibitem
\bibitem[G{\"o}tze(2009)]{GotzeBook}
W.~G{\"o}tze, \emph{Complex Dynamics of Glass-Forming Liquids -- A
  Mode-Coupling Theory}, Oxford University, Oxford, 2009\relax
\mciteBstWouldAddEndPuncttrue
\mciteSetBstMidEndSepPunct{\mcitedefaultmidpunct}
{\mcitedefaultendpunct}{\mcitedefaultseppunct}\relax
\EndOfBibitem
\bibitem[Foffi \emph{et~al.}(2002)Foffi, Dawson, Buldyrev, Sciortino,
  Zaccarelli, and Tartaglia]{FofDawBulSciZacTar02PRE}
G.~Foffi, K.~A. Dawson, S.~V. Buldyrev, F.~Sciortino, E.~Zaccarelli and
  P.~Tartaglia, \emph{Phys. Rev. E}, 2002, \textbf{65}, 050802\relax
\mciteBstWouldAddEndPuncttrue
\mciteSetBstMidEndSepPunct{\mcitedefaultmidpunct}
{\mcitedefaultendpunct}{\mcitedefaultseppunct}\relax
\EndOfBibitem
\bibitem[Zaccarelli \emph{et~al.}(2002)Zaccarelli, Foffi, Dawson, Buldyrev,
  Sciortino, and Tartaglia]{ZacFofDawBulSciTar02PRE}
E.~Zaccarelli, G.~Foffi, K.~A. Dawson, S.~V. Buldyrev, F.~Sciortino and
  P.~Tartaglia, \emph{Phys. Rev. E}, 2002, \textbf{66}, 041402\relax
\mciteBstWouldAddEndPuncttrue
\mciteSetBstMidEndSepPunct{\mcitedefaultmidpunct}
{\mcitedefaultendpunct}{\mcitedefaultseppunct}\relax
\EndOfBibitem
\bibitem[Sciortino \emph{et~al.}(2003)Sciortino, Tartaglia, and
  Zaccarelli]{SciTarZac03PRL}
F.~Sciortino, P.~Tartaglia and E.~Zaccarelli, \emph{Phys. Rev. Lett.}, 2003,
  \textbf{91}, 268301\relax
\mciteBstWouldAddEndPuncttrue
\mciteSetBstMidEndSepPunct{\mcitedefaultmidpunct}
{\mcitedefaultendpunct}{\mcitedefaultseppunct}\relax
\EndOfBibitem
\bibitem[Pham \emph{et~al.}(2004)Pham, Egelhaaf, Pusey, and
  Poon]{PhaEgePusPoo04PRE}
K.~N. Pham, S.~U. Egelhaaf, P.~N. Pusey and W.~C.~K. Poon, \emph{Phys. Rev. E},
  2004, \textbf{69}, 011503\relax
\mciteBstWouldAddEndPuncttrue
\mciteSetBstMidEndSepPunct{\mcitedefaultmidpunct}
{\mcitedefaultendpunct}{\mcitedefaultseppunct}\relax
\EndOfBibitem
\bibitem[Lang \emph{et~al.}(2010)Lang, Bo\ifmmode~\mbox{\c{t}}\else
  \c{t}\fi{}an, Oettel, Hajnal, Franosch, and
  Schilling]{LanBotOetHajFraSch10PRL}
S.~Lang, V.~Bo\ifmmode~\mbox{\c{t}}\else \c{t}\fi{}an, M.~Oettel, D.~Hajnal,
  T.~Franosch and R.~Schilling, \emph{Phys. Rev. Lett.}, 2010, \textbf{105},
  125701\relax
\mciteBstWouldAddEndPuncttrue
\mciteSetBstMidEndSepPunct{\mcitedefaultmidpunct}
{\mcitedefaultendpunct}{\mcitedefaultseppunct}\relax
\EndOfBibitem
\bibitem[Mandal \emph{et~al.}(2014)Mandal, Lang, Gross, Oettel, Raabe,
  Franosch, and Varnik]{ManLanGroOetRaaFraVar14NC}
S.~Mandal, S.~Lang, M.~Gross, M.~Oettel, D.~Raabe, T.~Franosch and F.~Varnik,
  \emph{Nat. Commun.}, 2014, \textbf{5}, 4435\relax
\mciteBstWouldAddEndPuncttrue
\mciteSetBstMidEndSepPunct{\mcitedefaultmidpunct}
{\mcitedefaultendpunct}{\mcitedefaultseppunct}\relax
\EndOfBibitem
\bibitem[G{\"o}tze and Sj{\"o}gren(1992)]{GotSjo92RPP}
W.~G{\"o}tze and L.~Sj{\"o}gren, \emph{Rep. Prog. Phys.}, 1992, \textbf{55},
  241--376\relax
\mciteBstWouldAddEndPuncttrue
\mciteSetBstMidEndSepPunct{\mcitedefaultmidpunct}
{\mcitedefaultendpunct}{\mcitedefaultseppunct}\relax
\EndOfBibitem
\bibitem[G{\"o}tze(1999)]{Got99JPCM}
W.~G{\"o}tze, \emph{J. Phys.: Condens. Matter}, 1999, \textbf{11},
  A1--A45\relax
\mciteBstWouldAddEndPuncttrue
\mciteSetBstMidEndSepPunct{\mcitedefaultmidpunct}
{\mcitedefaultendpunct}{\mcitedefaultseppunct}\relax
\EndOfBibitem
\bibitem[G{\"o}tze \emph{et~al.}(1981)G{\"o}tze, Leutheusser, and
  Yip]{GotLeuYip81aPRA}
W.~G{\"o}tze, E.~Leutheusser and S.~Yip, \emph{Phys. Rev. A}, 1981,
  \textbf{23}, 2634--2643\relax
\mciteBstWouldAddEndPuncttrue
\mciteSetBstMidEndSepPunct{\mcitedefaultmidpunct}
{\mcitedefaultendpunct}{\mcitedefaultseppunct}\relax
\EndOfBibitem
\bibitem[G{\"o}tze \emph{et~al.}(1981)G{\"o}tze, Leutheusser, and
  Yip]{GotLeuYip81bPRA}
W.~G{\"o}tze, E.~Leutheusser and S.~Yip, \emph{Phys. Rev. A}, 1981,
  \textbf{24}, 1008--1015\relax
\mciteBstWouldAddEndPuncttrue
\mciteSetBstMidEndSepPunct{\mcitedefaultmidpunct}
{\mcitedefaultendpunct}{\mcitedefaultseppunct}\relax
\EndOfBibitem
\bibitem[Leutheusser(1983)]{Leu83aPRA}
E.~Leutheusser, \emph{Phys. Rev. A}, 1983, \textbf{28}, 2510--2517\relax
\mciteBstWouldAddEndPuncttrue
\mciteSetBstMidEndSepPunct{\mcitedefaultmidpunct}
{\mcitedefaultendpunct}{\mcitedefaultseppunct}\relax
\EndOfBibitem
\bibitem[Szamel(2004)]{Sza04EL}
G.~Szamel, \emph{Europhys. Lett.}, 2004, \textbf{65}, 498--504\relax
\mciteBstWouldAddEndPuncttrue
\mciteSetBstMidEndSepPunct{\mcitedefaultmidpunct}
{\mcitedefaultendpunct}{\mcitedefaultseppunct}\relax
\EndOfBibitem
\bibitem[Krakoviack(2005)]{Kra05PRL}
V.~Krakoviack, \emph{Phys. Rev. Lett.}, 2005, \textbf{94}, 065703\relax
\mciteBstWouldAddEndPuncttrue
\mciteSetBstMidEndSepPunct{\mcitedefaultmidpunct}
{\mcitedefaultendpunct}{\mcitedefaultseppunct}\relax
\EndOfBibitem
\bibitem[Krakoviack(2005)]{Kra05JPCM}
V.~Krakoviack, \emph{J. Phys.: Condens. Matter}, 2005, \textbf{17},
  S3565--S3570\relax
\mciteBstWouldAddEndPuncttrue
\mciteSetBstMidEndSepPunct{\mcitedefaultmidpunct}
{\mcitedefaultendpunct}{\mcitedefaultseppunct}\relax
\EndOfBibitem
\bibitem[Krakoviack(2007)]{Kra07PRE}
V.~Krakoviack, \emph{Phys. Rev. E}, 2007, \textbf{75}, 031503\relax
\mciteBstWouldAddEndPuncttrue
\mciteSetBstMidEndSepPunct{\mcitedefaultmidpunct}
{\mcitedefaultendpunct}{\mcitedefaultseppunct}\relax
\EndOfBibitem
\bibitem[Krakoviack(2009)]{Kra09PRE}
V.~Krakoviack, \emph{Phys. Rev. E}, 2009, \textbf{79}, 061501\relax
\mciteBstWouldAddEndPuncttrue
\mciteSetBstMidEndSepPunct{\mcitedefaultmidpunct}
{\mcitedefaultendpunct}{\mcitedefaultseppunct}\relax
\EndOfBibitem
\bibitem[Krakoviack(2011)]{Kra11PRE}
V.~Krakoviack, \emph{Phys. Rev. E}, 2011, \textbf{84}, 050501(R)\relax
\mciteBstWouldAddEndPuncttrue
\mciteSetBstMidEndSepPunct{\mcitedefaultmidpunct}
{\mcitedefaultendpunct}{\mcitedefaultseppunct}\relax
\EndOfBibitem
\bibitem[Kurzidim \emph{et~al.}(2009)Kurzidim, Coslovich, and
  Kahl]{KurCosKah09PRL}
J.~Kurzidim, D.~Coslovich and G.~Kahl, \emph{Phys. Rev. Lett.}, 2009,
  \textbf{103}, 138303\relax
\mciteBstWouldAddEndPuncttrue
\mciteSetBstMidEndSepPunct{\mcitedefaultmidpunct}
{\mcitedefaultendpunct}{\mcitedefaultseppunct}\relax
\EndOfBibitem
\bibitem[Kurzidim \emph{et~al.}(2010)Kurzidim, Coslovich, and
  Kahl]{KurCosKah10PRE}
J.~Kurzidim, D.~Coslovich and G.~Kahl, \emph{Phys. Rev. E}, 2010, \textbf{82},
  041505\relax
\mciteBstWouldAddEndPuncttrue
\mciteSetBstMidEndSepPunct{\mcitedefaultmidpunct}
{\mcitedefaultendpunct}{\mcitedefaultseppunct}\relax
\EndOfBibitem
\bibitem[Kurzidim \emph{et~al.}(2011)Kurzidim, Coslovich, and
  Kahl]{KurCosKah11JPCM}
J.~Kurzidim, D.~Coslovich and G.~Kahl, \emph{J. Phys.: Condens. Matter}, 2011,
  \textbf{23}, 234122\relax
\mciteBstWouldAddEndPuncttrue
\mciteSetBstMidEndSepPunct{\mcitedefaultmidpunct}
{\mcitedefaultendpunct}{\mcitedefaultseppunct}\relax
\EndOfBibitem
\bibitem[Kim \emph{et~al.}(2009)Kim, Miyazaki, and Saito]{KimMiySai09EL}
K.~Kim, K.~Miyazaki and S.~Saito, \emph{EPL}, 2009, \textbf{88}, 36002\relax
\mciteBstWouldAddEndPuncttrue
\mciteSetBstMidEndSepPunct{\mcitedefaultmidpunct}
{\mcitedefaultendpunct}{\mcitedefaultseppunct}\relax
\EndOfBibitem
\bibitem[Kim \emph{et~al.}(2010)Kim, Miyazaki, and Saito]{KimMiySai10EPJST}
K.~Kim, K.~Miyazaki and S.~Saito, \emph{Eur. Phys. J. Special Topics}, 2010,
  \textbf{189}, 135--139\relax
\mciteBstWouldAddEndPuncttrue
\mciteSetBstMidEndSepPunct{\mcitedefaultmidpunct}
{\mcitedefaultendpunct}{\mcitedefaultseppunct}\relax
\EndOfBibitem
\bibitem[Kim \emph{et~al.}(2011)Kim, Miyazaki, and Saito]{KimMiySai11JPCM}
K.~Kim, K.~Miyazaki and S.~Saito, \emph{J. Phys.: Condens. Matter}, 2011,
  \textbf{23}, 234123\relax
\mciteBstWouldAddEndPuncttrue
\mciteSetBstMidEndSepPunct{\mcitedefaultmidpunct}
{\mcitedefaultendpunct}{\mcitedefaultseppunct}\relax
\EndOfBibitem
\bibitem[Spanner \emph{et~al.}(2013)Spanner, Schnyder, H{\"o}fling, Voigtmann,
  and Franosch]{SpaSchHofVoiFra13SM}
M.~Spanner, S.~K. Schnyder, F.~H{\"o}fling, T.~Voigtmann and T.~Franosch,
  \emph{Soft Matter}, 2013, \textbf{9}, 1604--1611\relax
\mciteBstWouldAddEndPuncttrue
\mciteSetBstMidEndSepPunct{\mcitedefaultmidpunct}
{\mcitedefaultendpunct}{\mcitedefaultseppunct}\relax
\EndOfBibitem
\bibitem[Alcoutlabi and Mc{K}enna(2005)]{AlcMcK05JPCM}
M.~Alcoutlabi and G.~B. Mc{K}enna, \emph{J. Phys.: Condens. Matter}, 2005,
  \textbf{17}, R461--R524\relax
\mciteBstWouldAddEndPuncttrue
\mciteSetBstMidEndSepPunct{\mcitedefaultmidpunct}
{\mcitedefaultendpunct}{\mcitedefaultseppunct}\relax
\EndOfBibitem
\bibitem[Alba-Simionesco \emph{et~al.}(2006)Alba-Simionesco, Coasne, Dosseh,
  Dudziak, Gubbins, Radhakrishnan, and
  Sliwinska-Bartkowiak]{AlbCoaDosDudGubRadSli06JPCM}
C.~Alba-Simionesco, B.~Coasne, G.~Dosseh, G.~Dudziak, K.~E. Gubbins,
  R.~Radhakrishnan and M.~Sliwinska-Bartkowiak, \emph{J. Phys.: Condens.
  Matter}, 2006, \textbf{18}, R15--R68\relax
\mciteBstWouldAddEndPuncttrue
\mciteSetBstMidEndSepPunct{\mcitedefaultmidpunct}
{\mcitedefaultendpunct}{\mcitedefaultseppunct}\relax
\EndOfBibitem
\bibitem[Richert(2011)]{Ric11ARPC}
R.~Richert, \emph{Annu. Rev. Phys. Chem.}, 2011, \textbf{62}, 65--84\relax
\mciteBstWouldAddEndPuncttrue
\mciteSetBstMidEndSepPunct{\mcitedefaultmidpunct}
{\mcitedefaultendpunct}{\mcitedefaultseppunct}\relax
\EndOfBibitem
\bibitem[Spanner \emph{et~al.}(2016)Spanner, H{\"o}fling, Kapfer, Mecke,
  Schr{\"o}der-Turk, and Franosch]{SpaHofKapMecSchFra16PRL}
M.~Spanner, F.~H{\"o}fling, S.~C. Kapfer, K.~R. Mecke, G.~E. Schr{\"o}der-Turk
  and T.~Franosch, \emph{Phys. Rev. Lett.}, 2016, \textbf{116}, 060601\relax
\mciteBstWouldAddEndPuncttrue
\mciteSetBstMidEndSepPunct{\mcitedefaultmidpunct}
{\mcitedefaultendpunct}{\mcitedefaultseppunct}\relax
\EndOfBibitem
\bibitem[Menon and Dasgupta(1994)]{MenDas94PRL}
G.~I. Menon and C.~Dasgupta, \emph{Phys. Rev. Lett.}, 1994, \textbf{73},
  1023--1026\relax
\mciteBstWouldAddEndPuncttrue
\mciteSetBstMidEndSepPunct{\mcitedefaultmidpunct}
{\mcitedefaultendpunct}{\mcitedefaultseppunct}\relax
\EndOfBibitem
\bibitem[Thalmann \emph{et~al.}(2000)Thalmann, Dasgupta, and
  Feinberg]{ThaDasFei00EL}
F.~Thalmann, C.~Dasgupta and D.~Feinberg, \emph{Europhys. Lett.}, 2000,
  \textbf{50}, 54--60\relax
\mciteBstWouldAddEndPuncttrue
\mciteSetBstMidEndSepPunct{\mcitedefaultmidpunct}
{\mcitedefaultendpunct}{\mcitedefaultseppunct}\relax
\EndOfBibitem
\bibitem[Kraichnan(1976)]{Kra76JFM}
R.~H. Kraichnan, \emph{J. Fluid Mech.}, 1976, \textbf{77}, 753--768\relax
\mciteBstWouldAddEndPuncttrue
\mciteSetBstMidEndSepPunct{\mcitedefaultmidpunct}
{\mcitedefaultendpunct}{\mcitedefaultseppunct}\relax
\EndOfBibitem
\bibitem[Lifshits \emph{et~al.}(1988)Lifshits, Gredeskul, and
  Pastur]{LifGrePasbook}
I.~M. Lifshits, S.~A. Gredeskul and L.~A. Pastur, \emph{Introduction to the
  theory of disordered systems}, Wiley, New York, 1988\relax
\mciteBstWouldAddEndPuncttrue
\mciteSetBstMidEndSepPunct{\mcitedefaultmidpunct}
{\mcitedefaultendpunct}{\mcitedefaultseppunct}\relax
\EndOfBibitem
\bibitem[Chudnovsky and Dickman(1998)]{ChuDic98PRB}
E.~M. Chudnovsky and R.~Dickman, \emph{Phys. Rev. B}, 1998, \textbf{57},
  2724--2727\relax
\mciteBstWouldAddEndPuncttrue
\mciteSetBstMidEndSepPunct{\mcitedefaultmidpunct}
{\mcitedefaultendpunct}{\mcitedefaultseppunct}\relax
\EndOfBibitem
\bibitem[G{\"o}tze(1978)]{Got78SSC}
W.~G{\"o}tze, \emph{Solid State Commun.}, 1978, \textbf{27}, 1393--1395\relax
\mciteBstWouldAddEndPuncttrue
\mciteSetBstMidEndSepPunct{\mcitedefaultmidpunct}
{\mcitedefaultendpunct}{\mcitedefaultseppunct}\relax
\EndOfBibitem
\bibitem[G{\"o}tze(1979)]{Got79JPC}
W.~G{\"o}tze, \emph{J. Phys. C: Solid State Phys.}, 1979, \textbf{12},
  1279--1296\relax
\mciteBstWouldAddEndPuncttrue
\mciteSetBstMidEndSepPunct{\mcitedefaultmidpunct}
{\mcitedefaultendpunct}{\mcitedefaultseppunct}\relax
\EndOfBibitem
\bibitem[G{\"o}tze \emph{et~al.}(1979)G{\"o}tze, Prelov{\v{s}}ek, and
  W{\"o}lfle]{GotPreWol79SSC}
W.~G{\"o}tze, P.~Prelov{\v{s}}ek and P.~W{\"o}lfle, \emph{Solid State Commun.},
  1979, \textbf{30}, 369--373\relax
\mciteBstWouldAddEndPuncttrue
\mciteSetBstMidEndSepPunct{\mcitedefaultmidpunct}
{\mcitedefaultendpunct}{\mcitedefaultseppunct}\relax
\EndOfBibitem
\bibitem[G{\"o}tze(1981)]{Got81PMB}
W.~G{\"o}tze, \emph{Philos. Mag. B}, 1981, \textbf{43}, 219--250\relax
\mciteBstWouldAddEndPuncttrue
\mciteSetBstMidEndSepPunct{\mcitedefaultmidpunct}
{\mcitedefaultendpunct}{\mcitedefaultseppunct}\relax
\EndOfBibitem
\bibitem[Leutheusser(1983)]{Leu83bPRA}
E.~Leutheusser, \emph{Phys. Rev. A}, 1983, \textbf{28}, 1762--1773\relax
\mciteBstWouldAddEndPuncttrue
\mciteSetBstMidEndSepPunct{\mcitedefaultmidpunct}
{\mcitedefaultendpunct}{\mcitedefaultseppunct}\relax
\EndOfBibitem
\bibitem[Schnyder \emph{et~al.}(2011)Schnyder, H{\"o}fling, Franosch, and
  Voigtmann]{SchHofFraVoi11JPCM}
S.~K. Schnyder, F.~H{\"o}fling, T.~Franosch and T.~Voigtmann, \emph{J. Phys.:
  Condens. Matter}, 2011, \textbf{23}, 234121\relax
\mciteBstWouldAddEndPuncttrue
\mciteSetBstMidEndSepPunct{\mcitedefaultmidpunct}
{\mcitedefaultendpunct}{\mcitedefaultseppunct}\relax
\EndOfBibitem
\bibitem[Hansen and McDonald(1986)]{macdohansen2ed}
J.-P. Hansen and I.~R. McDonald, \emph{Theory of simple liquids, Second
  edition}, Academic Press, London, 1986\relax
\mciteBstWouldAddEndPuncttrue
\mciteSetBstMidEndSepPunct{\mcitedefaultmidpunct}
{\mcitedefaultendpunct}{\mcitedefaultseppunct}\relax
\EndOfBibitem
\bibitem[Grinstein \emph{et~al.}(1977)Grinstein, Ma, and
  Mazenko]{GriMaMaz77PRB}
G.~Grinstein, S.-K. Ma and G.~F. Mazenko, \emph{Phys. Rev. B}, 1977,
  \textbf{15}, 258--272\relax
\mciteBstWouldAddEndPuncttrue
\mciteSetBstMidEndSepPunct{\mcitedefaultmidpunct}
{\mcitedefaultendpunct}{\mcitedefaultseppunct}\relax
\EndOfBibitem
\bibitem[Edwards and Anderson(1975)]{EdwAnd75JPF}
S.~F. Edwards and P.~W. Anderson, \emph{J. Phys. F: Metal Phys.}, 1975,
  \textbf{5}, 965--974\relax
\mciteBstWouldAddEndPuncttrue
\mciteSetBstMidEndSepPunct{\mcitedefaultmidpunct}
{\mcitedefaultendpunct}{\mcitedefaultseppunct}\relax
\EndOfBibitem
\bibitem[Deam and Edwards(1976)]{DeaEdw76PT}
R.~T. Deam and S.~F. Edwards, \emph{Philos. Trans. R. Soc. A}, 1976,
  \textbf{280}, 317--353\relax
\mciteBstWouldAddEndPuncttrue
\mciteSetBstMidEndSepPunct{\mcitedefaultmidpunct}
{\mcitedefaultendpunct}{\mcitedefaultseppunct}\relax
\EndOfBibitem
\bibitem[Lang \emph{et~al.}(2000)Lang, Likos, Watzlawek, and
  L{\"o}wen]{LanLikWatLow00JPCM}
A.~Lang, C.~N. Likos, M.~Watzlawek and H.~L{\"o}wen, \emph{J. Phys.: Condens.
  Matter}, 2000, \textbf{12}, 5087\relax
\mciteBstWouldAddEndPuncttrue
\mciteSetBstMidEndSepPunct{\mcitedefaultmidpunct}
{\mcitedefaultendpunct}{\mcitedefaultseppunct}\relax
\EndOfBibitem
\bibitem[Louis \emph{et~al.}(2000)Louis, Bolhuis, and Hansen]{LouBolHan00PRE}
A.~A. Louis, P.~G. Bolhuis and J.~P. Hansen, \emph{Phys. Rev. E}, 2000,
  \textbf{62}, 7961--7972\relax
\mciteBstWouldAddEndPuncttrue
\mciteSetBstMidEndSepPunct{\mcitedefaultmidpunct}
{\mcitedefaultendpunct}{\mcitedefaultseppunct}\relax
\EndOfBibitem
\bibitem[Krakoviack \emph{et~al.}(2003)Krakoviack, Hansen, and
  Louis]{KraHanLou03PRE}
V.~Krakoviack, J.-P. Hansen and A.~A. Louis, \emph{Phys. Rev. E}, 2003,
  \textbf{67}, 041801\relax
\mciteBstWouldAddEndPuncttrue
\mciteSetBstMidEndSepPunct{\mcitedefaultmidpunct}
{\mcitedefaultendpunct}{\mcitedefaultseppunct}\relax
\EndOfBibitem
\bibitem[Franosch \emph{et~al.}(1997)Franosch, Fuchs, G{\"o}tze, Mayr, and
  Singh]{FraFucGotMaySin97PRE}
T.~Franosch, M.~Fuchs, W.~G{\"o}tze, M.~R. Mayr and A.~P. Singh, \emph{Phys.
  Rev. E}, 1997, \textbf{55}, 7153--7176\relax
\mciteBstWouldAddEndPuncttrue
\mciteSetBstMidEndSepPunct{\mcitedefaultmidpunct}
{\mcitedefaultendpunct}{\mcitedefaultseppunct}\relax
\EndOfBibitem
\bibitem[Fuchs \emph{et~al.}(1998)Fuchs, G{\"o}tze, and Mayr]{FucGotMay98PRE}
M.~Fuchs, W.~G{\"o}tze and M.~R. Mayr, \emph{Phys. Rev. E}, 1998, \textbf{58},
  3384--3399\relax
\mciteBstWouldAddEndPuncttrue
\mciteSetBstMidEndSepPunct{\mcitedefaultmidpunct}
{\mcitedefaultendpunct}{\mcitedefaultseppunct}\relax
\EndOfBibitem
\bibitem[Fuchs \emph{et~al.}(1991)Fuchs, G{\"o}tze, Hofacker, and
  Latz]{FucGotHofLat91JPCM}
M.~Fuchs, W.~G{\"o}tze, I.~Hofacker and A.~Latz, \emph{J. Phys.: Condens.
  Matter}, 1991, \textbf{3}, 5047--5071\relax
\mciteBstWouldAddEndPuncttrue
\mciteSetBstMidEndSepPunct{\mcitedefaultmidpunct}
{\mcitedefaultendpunct}{\mcitedefaultseppunct}\relax
\EndOfBibitem
\bibitem[Franosch and G{\"o}tze(1994)]{FraGot94JPCM}
T.~Franosch and W.~G{\"o}tze, \emph{J. Phys.: Condens. Matter}, 1994,
  \textbf{6}, 4807--4822\relax
\mciteBstWouldAddEndPuncttrue
\mciteSetBstMidEndSepPunct{\mcitedefaultmidpunct}
{\mcitedefaultendpunct}{\mcitedefaultseppunct}\relax
\EndOfBibitem
\bibitem[Kert{\'e}sz and Metzger(1983)]{KerMet83JPA}
J.~Kert{\'e}sz and J.~Metzger, \emph{J. Phys. A: Math. Gen.}, 1983,
  \textbf{16}, L735--L739\relax
\mciteBstWouldAddEndPuncttrue
\mciteSetBstMidEndSepPunct{\mcitedefaultmidpunct}
{\mcitedefaultendpunct}{\mcitedefaultseppunct}\relax
\EndOfBibitem
\bibitem[Kammerer \emph{et~al.}(2008)Kammerer, H{\"o}fling, and
  Franosch]{KamHofFra08EL}
A.~Kammerer, F.~H{\"o}fling and T.~Franosch, \emph{EPL}, 2008, \textbf{84},
  66002\relax
\mciteBstWouldAddEndPuncttrue
\mciteSetBstMidEndSepPunct{\mcitedefaultmidpunct}
{\mcitedefaultendpunct}{\mcitedefaultseppunct}\relax
\EndOfBibitem
\bibitem[Franosch \emph{et~al.}(2011)Franosch, Spanner, Bauer,
  Schr{\"o}der-Turk, and H{\"o}fling]{FraSpaBauSchHof11JNCS}
T.~Franosch, M.~Spanner, T.~Bauer, G.~E. Schr{\"o}der-Turk and F.~H{\"o}fling,
  \emph{J. Non-Cryst. Solids}, 2011, \textbf{357}, 472--478\relax
\mciteBstWouldAddEndPuncttrue
\mciteSetBstMidEndSepPunct{\mcitedefaultmidpunct}
{\mcitedefaultendpunct}{\mcitedefaultseppunct}\relax
\EndOfBibitem
\bibitem[Spanner \emph{et~al.}(2011)Spanner, H{\"o}fling, Schr{\"o}der-Turk,
  Mecke, and Franosch]{SpaHofSchMecFra11JPCM}
M.~Spanner, F.~H{\"o}fling, G.~E. Schr{\"o}der-Turk, K.~Mecke and T.~Franosch,
  \emph{J. Phys.: Condens. Matter}, 2011, \textbf{23}, 234120\relax
\mciteBstWouldAddEndPuncttrue
\mciteSetBstMidEndSepPunct{\mcitedefaultmidpunct}
{\mcitedefaultendpunct}{\mcitedefaultseppunct}\relax
\EndOfBibitem
\bibitem[Franosch(2014)]{Fra14JPA}
T.~Franosch, \emph{J. Phys. A: Math. Theor.}, 2014, \textbf{47}, 325004\relax
\mciteBstWouldAddEndPuncttrue
\mciteSetBstMidEndSepPunct{\mcitedefaultmidpunct}
{\mcitedefaultendpunct}{\mcitedefaultseppunct}\relax
\EndOfBibitem
\bibitem[Bosse and Thakur(1987)]{BosTha87PRL}
J.~Bosse and J.~S. Thakur, \emph{Phys. Rev. Lett.}, 1987, \textbf{59},
  998\relax
\mciteBstWouldAddEndPuncttrue
\mciteSetBstMidEndSepPunct{\mcitedefaultmidpunct}
{\mcitedefaultendpunct}{\mcitedefaultseppunct}\relax
\EndOfBibitem
\bibitem[Thakur and Bosse(1991)]{ThaBos91PRA}
J.~S. Thakur and J.~Bosse, \emph{Phys. Rev. A}, 1991, \textbf{43},
  4378--4387\relax
\mciteBstWouldAddEndPuncttrue
\mciteSetBstMidEndSepPunct{\mcitedefaultmidpunct}
{\mcitedefaultendpunct}{\mcitedefaultseppunct}\relax
\EndOfBibitem
\bibitem[Thakur and Bosse(1991)]{ThaBos91PRAb}
J.~S. Thakur and J.~Bosse, \emph{Phys. Rev. A}, 1991, \textbf{43},
  4388--4395\relax
\mciteBstWouldAddEndPuncttrue
\mciteSetBstMidEndSepPunct{\mcitedefaultmidpunct}
{\mcitedefaultendpunct}{\mcitedefaultseppunct}\relax
\EndOfBibitem
\bibitem[Voigtmann(2011)]{Voi11EPL}
T.~Voigtmann, \emph{EPL}, 2011, \textbf{96}, 36006\relax
\mciteBstWouldAddEndPuncttrue
\mciteSetBstMidEndSepPunct{\mcitedefaultmidpunct}
{\mcitedefaultendpunct}{\mcitedefaultseppunct}\relax
\EndOfBibitem
\bibitem[Boon and Yip(1991)]{boonyip}
J.-P. Boon and S.~Yip, \emph{Molecular hydrodynamics}, Dover, New York,
  1991\relax
\mciteBstWouldAddEndPuncttrue
\mciteSetBstMidEndSepPunct{\mcitedefaultmidpunct}
{\mcitedefaultendpunct}{\mcitedefaultseppunct}\relax
\EndOfBibitem
\bibitem[Jepsen(1965)]{Jep65JMP}
D.~W. Jepsen, \emph{J. Math. Phys.}, 1965, \textbf{6}, 405--413\relax
\mciteBstWouldAddEndPuncttrue
\mciteSetBstMidEndSepPunct{\mcitedefaultmidpunct}
{\mcitedefaultendpunct}{\mcitedefaultseppunct}\relax
\EndOfBibitem
\bibitem[Dieterich \emph{et~al.}(1980)Dieterich, Fulde, and
  Peschel]{DieFulPes80AP}
W.~Dieterich, P.~Fulde and I.~Peschel, \emph{Adv. Phys.}, 1980, \textbf{29},
  527--605\relax
\mciteBstWouldAddEndPuncttrue
\mciteSetBstMidEndSepPunct{\mcitedefaultmidpunct}
{\mcitedefaultendpunct}{\mcitedefaultseppunct}\relax
\EndOfBibitem
\bibitem[Kutner(1981)]{Kut81PLA}
R.~Kutner, \emph{Phys. Lett. A}, 1981, \textbf{81}, 239--240\relax
\mciteBstWouldAddEndPuncttrue
\mciteSetBstMidEndSepPunct{\mcitedefaultmidpunct}
{\mcitedefaultendpunct}{\mcitedefaultseppunct}\relax
\EndOfBibitem
\bibitem[Kehr \emph{et~al.}(1981)Kehr, Kutner, and Binder]{KehKutBin81PRB}
K.~W. Kehr, R.~Kutner and K.~Binder, \emph{Phys. Rev. B}, 1981, \textbf{23},
  4931--4945\relax
\mciteBstWouldAddEndPuncttrue
\mciteSetBstMidEndSepPunct{\mcitedefaultmidpunct}
{\mcitedefaultendpunct}{\mcitedefaultseppunct}\relax
\EndOfBibitem
\end{mcitethebibliography}
%\bibliographystyle{rsc} %the RSC's .bst file

\providecommand*{\mcitethebibliography}{\thebibliography}
\csname @ifundefined\endcsname{endmcitethebibliography}
{\let\endmcitethebibliography\endthebibliography}{}

\end{document}